\documentclass[aps,prl,superscriptaddress,nofootinbib,floatfix]{revtex4-2}
\usepackage{amsfonts}
\usepackage{epsfig}
\usepackage{bm}
\usepackage{amssymb}
\usepackage{amsmath}
\usepackage{color}
\usepackage{subfigure}

\begin{document}

\title{Stochastic fluctuations and the relaxation time in transient relativistic fluids}

\date{\today}
\author{Gabriel S. Denicol}
\email{gsdenicol@id.uff.br}
\affiliation{Instituto de F\'{\i}sica, Universidade Federal Fluminense, Niter\'{o}i, Rio de Janeiro, 24210-346, Brazil}

\author{Jorge Noronha}
\email{jn0508@illinois.edu}
\affiliation{
Illinois Center for Advanced Studies of the Universe \& Department of Physics,\\
University of Illinois Urbana-Champaign, Urbana, IL 61801, USA
}

\begin{abstract}
We argue that the ratio between the shear viscosity and the shear relaxation time, $\eta/\tau_\pi$, should be defined as a thermodynamic quantity obtained from the equal-time symmetric correlator of the shear-stress tensor. In kinetic theory, we show that this ratio does not depend on the type of interaction. Similarly, an exact expression for this ratio is obtained for holographic gauge theories. We also determine how stochastic fluctuations change $\eta/\tau_\pi$ in transient relativistic hydrodynamics and show that thermal fluctuations do not spoil causality and stability.  

\end{abstract}

\maketitle

\affiliation{Instituto de F\'isica, Universidade Federal Fluminense, UFF, Niter\'oi,
24210-346, RJ, Brazil}

\affiliation{Illinois Center for Advanced Studies of the Universe\\
Department of Physics, University of Illinois at Urbana-Champaign, Urbana,
IL 61801, USA}



\section{Introduction}

Relativistic fluid dynamics is widely applied in high-energy nuclear physics to describe the hot and dense nuclear matter produced in ultra-relativistic heavy ion collisions \cite{Gale:2013da,Heinz:2013th,Romatschke:2017ejr} as well as the highly compressed baryonic matter that is produced in neutron star mergers \cite{Baiotti:2016qnr}. These studies aim to understand how the thermodynamic and transport properties of QCD matter change when subjected to extreme temperatures, pressures, and external fields (e.g., gravitational and electromagnetic fields). Such investigations have also considerably improved our understanding of relativistic fluid dynamics itself \cite{Rocha:2023ilf}, which can display qualitative differences from its non-relativistic counterpart, the traditional Navier-Stokes theory \cite{LandauLifshitzFluids}. In fact, the generalization of the Navier-Stokes equations to the relativistic domain proposed by Eckart \cite{EckartViscous} and Landau and Lifshitz \cite{LandauLifshitzFluids} display acausal behavior \cite{PichonViscous} which, when combined with dissipation, invariably implies that its global equilibrium state is unstable against small perturbations \cite{Hiscock_Lindblom_instability_1985,Denicol:2008ha, Pu:2009fj}, a general result explained in \cite{Gavassino:2021owo}.

Fluid-dynamical theories for relativistic fluids that can be linearly causal and stable when perturbed around global equilibrium were first proposed by Israel and Stewart in the 1970s \cite{Israel:1976tn, MIS-6}. In this type of theory, the dissipative currents such as the shear-stress tensor, $\pi_{\mu\nu}$, are not determined by constitutive relations, which to first order in derivatives set $\pi_{\mu\nu} = 2 \eta \sigma_{\mu\nu}$ \cite{EckartViscous,LandauLifshitzFluids}, where $\eta$ is the shear viscosity and $\sigma_{\mu\nu}$ is the shear tensor. Instead, the dissipative currents satisfy additional equations of motion that describe how they may relax to their corresponding Navier-Stokes-like constitutive relations. A novel set of transport coefficients determines the time scale over which this relaxation process occurs: the relaxation times\footnote{In the BDNK formalism \cite{Bemfica:2017wps,Kovtun:2019hdm,Bemfica:2019knx,Hoult:2020eho,Bemfica:2020zjp}, relaxation time coefficients may be defined in terms of hydrodynamic frame dependent coefficients, which effectively parametrize the non-hydrodynamic sector and our ignorance about the UV properties of the underlying microscopic theory.}. Neglecting other dissipative sources but shear viscosity, causality and stability in the linear regime around equilibrium require that the corresponding shear relaxation time $\tau_\pi$ is non-negative and cannot be arbitrarily small compared to the viscous attenuation length $\eta/(\varepsilon+P)$ \cite{Hiscock_Lindblom_stability_1983}, where $\varepsilon$ is the rest-frame energy density and $P$ is the thermodynamic pressure. 

Relaxation times thus play an essential role in relativistic viscous fluid-dynamical frameworks \cite{Rocha:2023ilf} and their applications \cite{Gale:2013da}. 
Nevertheless, the microscopic origin and interpretation of the relaxation time in relativistic fluids are still somewhat controversial \cite{Wagner:2023jgq}. As a matter of fact, while the shear viscosity is determined unambiguously from the retarded Green's function via Kubo formulas \cite{Kovtun:2012rj}, the relaxation time in causal and stable theories can still be interpreted in several different forms. The microscopic origin of the relaxation time was discussed in Ref.\ \cite{Denicol:2011fa} where it was proposed that $\tau_\pi$ should be generally determined by the slowest, purely imaginary and stable \emph{non-hydrodynamic} pole of retarded correlators in Fourier space. This interpretation was motivated by general properties of kinetic systems \cite{Denicol:2011fa} and later played a role in the development of the DNMR formalism \cite{Denicol:2012cn}. Another common interpretation advocated in \cite{Baier:2007ix} is that $\tau_\pi$ can be obtained from a Kubo formula derived from the series expansion of retarded correlators in the infrared.
In general, these different prescriptions do not give the same value for the relaxation time coefficient \cite{Denicol:2011fa}. Furthermore, both approaches have relevant shortcomings. 

Determining $\tau_\pi$ from the slowest non-hydrodynamic pole \cite{Denicol:2011fa} makes sense, in principle, if the non-hydrodynamic sector is strictly not gapless. However, there are examples where the correlator of the shear-stress tensor has a branch cut extending all the way down to zero frequency, such as in weakly coupled $\lambda \phi^4$ theory \cite{Moore:2018mma,Ochsenfeld:2023wxz,Rocha:2024cge}. In fact, it has been recently proven in \cite{Gavassino:2024rck} that the non-hydrodynamic sector is gapless in any relativistic kinetic theory whose scattering cross-section vanishes at large energies. On the other hand, the determination of $\tau_\pi$ from a Kubo formula via a gradient expansion \cite{Baier:2007ix,Moore:2010bu,Czajka:2017bod} does not exclude cases where $\tau_\pi <0$ \cite{Schaefer:2014aia}, which is problematic from the point of view of causality and stability of the corresponding hydrodynamic theory. Additionally, it is known that thermal fluctuations generate an $\omega^{3/2}$ term (where $\omega$ is the frequency) in the retarded Green's function in Fourier space \cite{DESCHEPPER19741,Kovtun:2003vj,Kovtun:2011np,Akamatsu:2016llw}, which makes the series nonanalytic at $\omega = 0$ and formally implies that the $\tau_\pi$ defined via a Kubo formula diverges \cite{Kovtun:2011np}. This was interpreted at the time as signaling the breakdown of second-order relativistic hydrodynamics  \cite{Kovtun:2011np}. Another consequence of this finding is that the fate of causality and stability in stochastic relativistic fluids was left hanging in the balance, as the very parameter that controlled such properties apparently became ill-defined due to the inexorable backreaction of the fluctuations present in a thermal medium.     

Motivated by the discussion above and other recent works \cite{Wagner:2023jgq,Mullins:2023tjg,SoaresRocha:2024afv}, in this paper, we reassess the microscopic interpretation of the relaxation time in transient relativistic fluid dynamics. We show how $\eta/\tau_\pi$ can be defined as a thermodynamic quantity given by the equal-time correlator of the shear-stress tensor computed in linear response. This prescription for $\eta/\tau_\pi$ can remain meaningful even if the non-hydrodynamic sector is gapless. In addition, $\eta/\tau_\pi$ defined in this form acquires purely thermodynamic meaning, being thus directly computable at weak coupling using kinetic theory, strong coupling using holography, and also, in principle, on the lattice. We further determine the backreaction of thermal fluctuations on $\eta/\tau_\pi$ and $\eta$ to find how the $\tau_\pi$ defined using our new prescription is changed at one loop order. Since our prescription differs from the one used in \cite{Kovtun:2011np}, $\tau_\pi$ remains meaningful in the presence of fluctuations even when $\omega \to 0$. Finally, we show that thermal fluctuations preserve the causality and stability lower bound on $\eta/\tau_\pi$.

\underline{Definitions:} We use a mostly minus Minkowski
metric $\eta _{\mu \nu }$ with coordinates $x^{\mu }=(t,\mathbf{x})$ and
natural units $\hbar =c=k_{B}=1$. The Fourier transform of a field, $A(x)$,
is always depicted as $\Tilde{A}(q)$. We adopt the following convention for
the Fourier transform, 
\begin{eqnarray}
\Tilde{A}(q) &=&\int d^{4}x\exp {(-iq_{\mu }x^{\mu })}A(x), \\
A(x) &=&\int \frac{d^{4}q}{(2\pi )^{4}}\exp {(iq_{\mu }x^{\mu })}\Tilde{A}%
(q).
\end{eqnarray}
The projection operator onto the subspace orthogonal to a unitary 4-vector $%
u^{\mu }$ is defined as, $\Delta ^{\mu \nu }=g^{\mu \nu }-u^{\mu }u^{\nu }$. The double, traceless, and symmetric projection operator is defined as, 
\begin{equation}
\Delta _{\alpha \beta }^{\mu \nu }=\frac{1}{2}\left( \Delta _{\alpha }^{\mu
}\Delta _{\beta }^{\nu }+\Delta _{\beta }^{\mu }\Delta _{\alpha }^{\mu }-%
\frac{2}{3}\Delta ^{\mu \nu }\Delta _{\alpha \beta }\right) .
\end{equation}%
Finally, the following notation is used for the traceless and symmetric
projection of a rank-2 tensor, $A^{\left\langle \mu \nu \right\rangle }\equiv \Delta _{\alpha \beta }^{\mu \nu }A^{\alpha\beta }$.

\section{New definition of the relaxation time}
\label{Hydro}

We consider a viscous fluid at zero chemical potential in 4-dimensional Minkowski spacetime with energy-momentum tensor \cite{LandauLifshitzFluids} 
\begin{equation}
T^{\mu \nu }=\varepsilon u^{\mu }u^{\mu }-P\Delta ^{\mu \nu }+\pi ^{\mu \nu
},  \label{conformalTmunu}
\end{equation}%
where $u^{\mu }$ is the 4-velocity ($%
u^{\mu }u_{\mu }=1$), and $\pi^{\mu\nu}u_\nu=0$ and $\pi^\mu_\mu=0$. We use the 
Landau frame \cite{LandauLifshitzFluids} and, for simplicity, only shear viscosity effects are considered.

The dynamics of the fluid is defined by the conservation of energy and
momentum, $\partial_{\mu }T^{\mu \nu }=0$, written here as 
\begin{subequations}
\begin{eqnarray}
&&u^{\mu }\partial_{\mu }\varepsilon +\left( \varepsilon +P\right) \partial_{\mu }u^{\mu
}+\pi ^{\mu \nu }\partial_{\mu }u_{\nu }=0  \label{defineEOMenergyIS} \\
&&\left( \varepsilon +P\right) u^{\lambda }\partial_{\lambda }u^{\nu }-\Delta ^{\nu
\lambda }\partial_{\lambda }P+\Delta _{\lambda }^{\nu }\partial_{\mu }\pi ^{\mu \lambda
}=0. \label{defineEOMmomentumIS}
\end{eqnarray}%
\end{subequations}
In transient fluid dynamics \cite{MIS-6}, the conservation laws are complemented with an additional hypothesis concerning the nature of
the shear-stress tensor. We follow Israel and
Stewart \cite{MIS-2,MIS-6} and assume that $\pi ^{\mu \nu }$ obeys a
relaxation equation 
\begin{equation}
\tau _{\pi }\Delta _{\alpha \beta }^{\mu \nu }u^{\lambda }\partial_{\lambda }\pi
^{\alpha \beta }+\pi ^{\mu \nu }-2\eta \sigma ^{\mu \nu }+\mathcal{O}^{\mu\nu} =0 , \label{defineISshear}
\end{equation}%
where $\sigma ^{\mu \nu }=\Delta^{\mu \nu\alpha \beta }\partial _{\alpha
}u_{\beta }$ is the shear tensor, and $\mathcal{O}^{\mu\nu}$ denotes higher-order nonlinear terms  \cite{Denicol:2012cn}, which do not contribute to linear response around equilibrium and are, thus, omitted.   Causality and stability in the linear regime
provides the following simple constraint for the relaxation time \cite%
{Hiscock_Lindblom_stability_1983,Olson:1989ey,Pu:2009fj,Brito:2020nou}, 
\begin{equation}
\tau _{\pi }\geq 2\frac{\eta }{\varepsilon +P}.  \label{causality}
\end{equation}%
In the nonlinear regime, where $O^{\mu\nu}$ can contribute, causality conditions were obtained in \cite{Bemfica:2020xym}.

To determine the relevant correlators, we assume the system is driven out of equilibrium by metric
perturbations, $g^{\mu \nu }=\eta ^{\mu \nu }+h^{\mu \nu }$, that in principle can lead to disturbances on all the fluid-dynamical fields,
$\{\delta \varepsilon,\delta u^{\mu},\delta \pi^{\mu \nu }\}$. 
Since we are solely interested in perturbations of the shear-stress tensor, we restrict ourselves to metric perturbations orthogonal to $u^{\mu }$ \textit{and} traceless: $u^{\mu }h_{\mu \nu
}=\eta ^{\mu \nu }h_{\mu \nu }=0$. Finally, we further assume that unperturbed fluid is homogeneous and isotropic.

For the sake of convenience, all our calculations will be
performed in Fourier space, in terms of the Fourier transformed fluid-dynamical perturbations $\{\delta \tilde{\varepsilon}(q),\delta \tilde{u}^{\mu }(q),\delta \tilde{\pi}^{\mu \nu }(q)\}$. We then decompose the Fourier 4-momentum $q^{\mu }$ in terms of $u^{\mu }$,
\begin{equation}
q^{\mu }=\Omega u^{\mu }+\mathfrak{q}^{\mu },
\end{equation}%
where $\Omega \equiv u_\mu q^\mu$ and $\mathfrak{q}^{\mu} \equiv \Delta^{\mu}_{\nu} q^{\nu}$. The linearized fluid-dynamical equations in Fourier space simplify dramatically,
\begin{eqnarray}
\Omega \delta \Tilde{\varepsilon} =0, \, \, \, \Omega \delta \Tilde{u}^{\mu
}&=&0, \\
i\Omega \tau_{\pi }\delta \Tilde{\pi}^{\mu \nu }+\delta \Tilde{\pi}^{\mu \nu }
&=&i\eta\Omega \Delta ^{\mu \nu \alpha \beta }\Tilde{h}_{\alpha \beta },
\end{eqnarray}
where we used that, up to first order in metric perturbations, $u_{\mu
}\delta u^{\mu }=0$ and $%
u_{\mu }\delta \pi ^{\mu \nu }=0$. As expected, we see from the equations above that only the fluctuations of the shear-stress tensor become nontrivial.

The fluctuations of the energy-momentum tensor, $\delta \Tilde{T}^{\mu \nu }$, can then be expressed as,
\begin{equation}
\delta \Tilde{T}^{\mu \nu }=\delta \Tilde{\pi}^{\mu \nu }-P\Tilde{h}^{\mu \nu }=\frac{i\Omega \eta 
}{1+i\Omega \tau _{\pi }}\Delta ^{\mu \nu \alpha \beta }\Tilde{h}_{\alpha \beta }-
\frac{P}{\varepsilon +P}\Tilde{h}^{\mu \nu }.
\end{equation}%
Thus, the energy-momentum tensor fluctuations can be cast in the following form, typical of linear response theory, 
\begin{equation}
\delta \tilde{T}^{\mu \nu }(q)=\frac{1}{2}\tilde{G}_{R}^{\mu \nu \alpha \beta}
(q)\tilde{h}_{\alpha \beta }(q),  \label{Linear_Response}
\end{equation}%
with the retarded Green's function, $\tilde{G}_{R}^{\mu \nu \alpha \beta}$, defined as implied. The imaginary part of the retarded Green's is then identified as
\begin{equation}
\mathcal{G}^{\mu \nu \alpha \beta }=2\Delta ^{\mu \nu \alpha \beta }\frac{\Omega \eta }{1+\Omega^2 \tau^2 _{\pi }}=\mathcal{G}\left( \Omega
\right) \Delta ^{\mu \nu \alpha \beta },
\end{equation}%
where the Lorentz scalar $\mathcal{G}\left( \Omega \right)$ is given by,
\begin{equation}
\mathcal{G}\left( \Omega \right) =\frac{2\Omega \eta }{1+\Omega ^{2}\tau
_{\pi }^{2}}. \label{IS_GreenFunction}
\end{equation}%

The transport properties of the fluid related to the shear-stress tensor are contained in the spectral density $\mathcal{G}$. The usual expectation is that the behavior of the retarded Green's function at low frequency will match those of the underlying microscopic theory and determine the fluid-dynamical properties of the system. For instance, $\eta $, is determined from the Kubo formula 
\begin{equation}
2\eta \equiv \lim_{\Omega \rightarrow 0}\frac{\mathcal{G}\left( \Omega
,0\right)}{\Omega}. \label{Kubo_eta}
\end{equation}

As discussed previously, there are different prescriptions to determine the relaxation time. For instance, the relaxation time was investigated using the projection operator method \cite{Koide:2009sy}.
In Ref.\ \cite{Denicol:2011fa}, the relaxation time is obtained from the pole of the retarded Green's function $G_R^{xyxy}$ closest to the origin. Another possibility is to expand the retarded Green's function near the origin and   
define the relaxation time in terms of second derivatives $G_R^{xyxy}$  \cite{Baier:2007ix,Moore:2010bu}. Naturally, in a fluid-dynamical framework, all these procedures lead to the same relaxation time. However, when applied to the retarded Green's function of the underlying microscopic theory, these procedures yield different results \cite{Denicol:2011fa}, with the setbacks discussed before. 

In this work, we consider another \emph{definition} for the relaxation time based on the following identity valid in transient hydrodynamics, see \eqref{IS_GreenFunction},
\begin{equation}
\frac{1}{\tau _{\pi }} \equiv \frac{1}{\eta }\int \frac{d\Omega }{2\pi }\frac{%
\mathcal{G}\left( \Omega ,0\right) }{\Omega }.
\label{prescription}
\end{equation}
This is analogous to the $f$-sum rule discussed in \cite{forster}. We demonstrate that $\tau_\pi$ defined in this way does not have the setbacks discussed above and can be consistently extended to stochastic hydrodynamics. The integral in \eqref{prescription} is a thermodynamic quantity, which
is compatible with the phenomenological derivation of Israel-Stewart theory where $\eta/\tau_\pi$ can be understood as a thermodynamic susceptibility associated with shear stress \cite{MIS-2,MIS-3,MIS-4,Gavassino:2021kjm,Mullins:2023tjg}.

The relaxation time defined in \eqref{prescription} can be calculated exactly in kinetic theory, where the retarded Green's function is obtained with linear response theory due to metric perturbation -- the details can be found in the Appendix. The integral in the right-hand side of \eqref{prescription} is then evaluated to be, 
\begin{equation}
\int \frac{d\Omega }{2\pi }\frac{%
\mathcal{G}\left( \Omega ,0\right) }{\Omega }=\frac{1 
}{10T}\int \frac{d^3k}{(2\pi)^3k^0}f_{0\mathbf{k}}(1+af_{0\mathbf{k}})E_{\mathbf{k}%
}^{-1}k_{\left\langle \mu\right. }k_{\left. \nu \right\rangle }
k^{\left\langle \mu \right. }k^{\left. \nu \right\rangle },  \label{Sum_Rule0}
\end{equation}%
where $k^\mu = (k^0,\mathbf{k})$ is the particle's 4-momentum, $E_{\mathbf{k}}=u_\mu k^\mu$, and  $f_{0\mathbf{k}}$ is the equilibrium distribution, with $a = +(-)$ for bosons (fermions), and $a = 0$ for classical statistics. This result is universal in kinetic theory and does not depend on the type of interactions. For massless particles, $\tau_\pi$ simplifies considerably and is universally given by,
\begin{equation}
    \tau_\pi = 5\frac{\eta}{\varepsilon+P},
\end{equation}
In particular, this holds for massless $\lambda \phi^4$ scalar theory, which possesses gapless non-hydrodynamic modes \cite{Moore:2018mma,Ochsenfeld:2023wxz,Rocha:2024cge,Gavassino:2024rck}. Coincidentally, this is also the expression traditionally derived from kinetic theory using the method of moments combined with the 14-moment approximation 
\cite{Denicol:2011fa,Denicol:2012cn}. It is interesting to note that the relaxation time calculated in kinetic theory for massless particles using the alternative prescription via the gradient expansion \cite{Baier:2007ix} is not universal, though it was shown to satisfy the inequality $\tau_{\pi} \geq 5\eta/(\varepsilon+P)$ \cite{Ghiglieri:2018dgf}.

We also note that the integral in \eqref{prescription} also appeared in the so-called shear spectral sum rule derived in \cite{Romatschke:2009ng} using linear response in the context of holographic strongly coupled gauge theories. In this case, vacuum contributions must be subtracted when calculating the spectral function, which is a nontrivial procedure in gauge theories; see also \cite{Meyer:2010gu}. The vacuum-subtracted result obtained in \cite{Romatschke:2009ng} is (using our notation), 
\begin{equation}
\int \frac{d\Omega }{2\pi }\frac{%
\mathcal{G}\left( \Omega ,0\right) }{\Omega }=\frac{2\varepsilon}{5}, \label{Sum_Rule1}
\end{equation}
which is valid for all strongly coupled gauge theories with a gravity dual for which $\eta/s = 1/4\pi$ \cite{Kovtun:2004de}. Thus, using the new definition of $\tau_\pi$ proposed here, one obtains the following universal relaxation time for this class of microscopic theories,
\begin{equation}
    \tau_\pi = \frac{10}{3} \frac{\eta}{\varepsilon+P} = \frac{10}{6}\frac{1}{2\pi T},
\end{equation}
which obeys the causality and stability bound. On the other hand, the relaxation time calculated using Kubo formulas derived within a gradient expansion is not universal in holography. In $N=4$ super Yang-Mills at infinite t'Hooft coupling and number of colors, one finds $\tau_\pi^{\mathrm{BRSSS}} = (2-\ln 2)/(2\pi T)$ \cite{Baier:2007ix}. Additionally, we also note that Ref.\ \cite{Meyer:2010gu} investigated the shear sum rule in pure glue gauge theory on the lattice, showing that, in principle, similar calculations of $\eta/\tau_\pi$ could be done in QCD.

Finally, the integrals of the spectral function in Eqs.~\eqref{Sum_Rule0} and \eqref{Sum_Rule1} 
diverge in Navier-Stokes theory. This implies that Navier-Stokes theory can never describe this thermodynamic susceptibility. The crucial question that one may pose is whether a hydrodynamic theory should be able to describe this feature. By introducing a relaxation time, 
we showed above that transient relativistic hydrodynamics can capture this property. This is a fundamental difference between these two hydrodynamic formulations.

\section{Effect of thermal fluctuations}

The backreaction of thermal fluctuations changes the spectral density $\mathcal{G}$ and, consequently, the values of $\eta$ and $\eta/\tau_\pi$. These corrections will be determined here using the symmetrized correlator of the shear stress tensor in transient relativistic hydrodynamics with noise \cite{Young:2014pka,Murase:2019cwc,Mullins:2023tjg}. 

We consider the linearized version of \eqref{defineEOMenergyIS}, \eqref{defineEOMmomentumIS}, and \eqref{defineISshear}  around a
global equilibrium state, with constant and uniform energy density, $\varepsilon $, and
4-velocity, $u^{\mu }$, and a vanishing shear stress tensor, $\pi ^{\mu \nu
}=0$. A stochastic source term, $S^{\mu\nu}$, leads to spacetime dependent fluctuations in the dynamical variables, $\delta \varepsilon (x)$, $\delta u^{\mu }(x)$, and $\delta \pi
^{\mu \nu }(x)$ which, to first order in the fluctuations, obey $u_{\mu
}\delta u^{\mu }(x)=0$ and $u_{\mu }\delta \pi ^{\mu \nu }(x)=0$. The Gaussian noise term satisfies $\langle S^{\mu \nu }(x)\rangle =0$, where $\langle \ldots
\rangle $ denotes the equilibrium average, and its 2-point correlation function is given by \cite{landau1980statisticalII}
\begin{equation}
\langle S^{\mu \nu }(x)S^{\alpha \beta }(x^{\prime })\rangle =4\eta T\Delta
^{\mu \nu \alpha \beta }\delta ^{(4)}(x-x^{\prime }),
\label{noisecorrelator}
\end{equation}
where $T$ is the background temperature.

The corresponding linearized versions of \eqref{defineEOMenergyIS}, \eqref{defineEOMmomentumIS}, and \eqref{defineISshear} in the presence of
noise define the following set of linear \emph{inhomogeneous} stochastic PDE's, 
\begin{subequations}
\begin{eqnarray}
&&u^{\mu }\partial _{\mu }\delta \varepsilon +\frac{4\varepsilon }{3}%
\partial _{\mu }\delta u^{\mu }=0  \label{defineISshearlinear} \\
&&\frac{4\varepsilon }{3}u^{\lambda }\partial _{\lambda }\delta u^{\nu }-%
\frac{1}{3}\Delta ^{\nu \lambda }\partial _{\lambda }\delta \varepsilon
+\Delta _{\lambda }^{\nu }\partial _{\mu }\delta \pi ^{\mu \lambda }=0
\label{defineISshearlinear2} \\
&&\tau _{\pi }u^{\lambda }\partial _{\lambda }\delta \pi ^{\mu \nu }+\delta
\pi ^{\mu \nu }-2\eta \delta \sigma ^{\mu \nu }=S^{\mu \nu }.
\label{ISlinear}
\end{eqnarray}%
\end{subequations}
To simplify the calculations, here we set the equation of state to be $P = \varepsilon/3$. 

The general solution for each of the stochastic fields $\{\delta \varepsilon
(x),\delta u^{\mu }(x),\delta \pi ^{\mu \nu }(x)\}$ is the sum of a
homogeneous part, which quickly decays with time\footnote{This only holds if the theory is linearly causal and stable.}, and a particular
solution that depends on the stochastic source, which then defines the long
time behavior of correlation functions. In this work, we are interested in
the long-time effects induced by fluctuations and, thus, we assume that $%
t/\tau_{\pi}\gg 1$, which allows us to safely keep only the particular
solutions in the following. Those can be most easily found in Fourier space,
as we explain below.

We now compute the \emph{connected}
symmetrized correlators of the energy-momentum tensor. In particular, we are
interested in the correlations in the shear channel, which emerge from the symmetric and traceless projection of $T^{\mu \nu }$,
\begin{equation}
T^{\left\langle \mu \nu \right\rangle }=\delta \pi ^{\mu \nu }+(\varepsilon+P)
\delta u^{\left\langle \mu \right. }\delta u^{\left. \nu \right\rangle }.
\end{equation}%
The relevant correlation function has the following general form,
\begin{equation}
G^{\mu\nu\alpha\beta} = \langle T^{\left\langle \mu \nu \right\rangle }\left( x\right) T^{\left\langle \alpha \beta \right\rangle }\left( x^{\prime }\right) \rangle -\langle T^{\left\langle \mu \nu \right\rangle }\left(
x\right)\rangle \langle T^{\left\langle \alpha \beta \right\rangle }\left( x^{\prime }\right) \rangle 
=G\left( x-x^{\prime }\right) \Delta^{\mu \nu \alpha \beta}.
\end{equation}
Above, $G(x-x')$ is a Lorentz scalar function that completely characterizes
the energy-momentum tensor correlations in the shear channel and can be determined from its
following projection,
\begin{equation}
G\left( x-x^{\prime }\right) = \frac{1}{5}\Delta _{\mu \nu \alpha \beta} G^{\mu\nu\alpha\beta}
= \frac{1}{5} \langle \delta \pi_{\mu \nu }\left( x\right) \delta \pi^{\mu \nu }\left( x^{\prime }\right) \rangle
+\frac{2}{5}(\varepsilon+P) ^{2}\Delta _{\mu \nu \alpha \beta }\langle
\delta u^{\mu }\left( x\right) \delta u^{\alpha }\left( x^{\prime }\right)
\rangle \langle \delta u^{\nu }\left( x\right) \delta u^{\beta
}\left( x^{\prime }\right) \rangle ,
\end{equation}
where we used the standard factorization $\langle A_{1}A_{2}A_{3}A_{4}\rangle
=\langle A_{1}A_{2}\rangle \langle A_{3}A_{4}\rangle +\langle
A_{1}A_{3}\rangle \langle A_{2}A_{4}\rangle +\langle A_{1}A_{4}\rangle
\langle A_{2}A_{3}\rangle$ to simplify the 4-point functions. The second term is a non-linear contribution to the correlation function in terms of 4-velocity correlators. The Fourier transform of the scalar correlation function is given by,
\begin{eqnarray}
\tilde{G}\left( q\right) \left( 2\pi \right) ^{4}\delta ^{4}\left(
q+q^{\prime }\right)  &=&\frac{1}{5}\langle \delta \pi_{\mu \nu }\left( q\right) \delta \pi^{\mu \nu }\left( q^{\prime }\right) \rangle   \notag \\
&+&\frac{2}{5}(\varepsilon+P)^{2}\Delta _{\mu \nu \alpha \beta }\int \frac{%
d^{4}pd^{4}p^{\prime }}{(2\pi )^{8}}\langle \delta \tilde{u}^{\mu
}\left( p\right) \delta \tilde{u}^{\alpha }\left( p^{\prime }\right)
\rangle \langle \delta \tilde{u}^{\nu }\left( q-p\right) \delta 
\tilde{u}^{\beta }\left( q^{\prime }-p^{\prime }\right) \rangle ,
\label{FullCorrelation}
\end{eqnarray}%
where we work with Fourier transformed fields $\{\delta \tilde{\varepsilon}%
(q),\delta \tilde{u}^{\mu }(q),\delta \tilde{\pi}^{\mu \nu }(q)\}$.

Equations \eqref{defineISshearlinear}-- \eqref{ISlinear} can be solved in Fourier space, leading to the following leading-order expressions for the shear-stress tensor and 4-velocity correlators,
\begin{eqnarray}
\langle \delta \tilde{\pi}^{\mu \nu }\left( q\right) \delta \tilde{\pi}%
^{\alpha \beta }\left( q^{\prime }\right) \rangle  &=&2\eta T\left[ \frac{4%
}{3}\left( \Omega ^{2}-\frac{\mathfrak{q}^{2}}{3}\right) ^{2}\left\vert G_{%
\text{sound}}(q)\right\vert ^{2}\left( \mathfrak{\hat{q}}^{\mu }\mathfrak{%
\hat{q}}^{\nu }+\frac{1}{2}\Delta _{\bot }^{\mu \nu }\right) \left( 
\mathfrak{\hat{q}}^{\alpha }\mathfrak{\hat{q}}^{\beta }+\frac{1}{2}\Delta
_{\bot }^{\alpha \beta }\right) +\frac{2\Delta _{\bot }^{\mu \nu \alpha
\beta }}{1+\tau _{\pi }^{2}\Omega ^{2}}\right.  \\
&&\left. -\Omega ^{2}\left\vert G_{\text{shear}}(q)\right\vert ^{2}\left(
\Delta _{\bot }^{\mu \alpha }\mathfrak{\hat{q}}^{\nu }\mathfrak{\hat{q}}%
^{\beta }+\Delta _{\bot }^{\mu \beta }\mathfrak{\hat{q}}^{\nu }\mathfrak{%
\hat{q}}^{\alpha }+\Delta _{\bot }^{\nu \alpha }\mathfrak{\hat{q}}^{\mu }%
\mathfrak{\hat{q}}^{\beta }+\Delta _{\bot }^{\nu \beta }\mathfrak{\hat{q}}%
^{\mu }\mathfrak{\hat{q}}^{\alpha }\right) \right] \left( 2\pi \right)
^{4}\delta ^{(4)}(q+q^{\prime }), \notag \\ 
\langle \delta \tilde{u}^{\mu }\left( q\right) \delta \tilde{u}^{\nu }\left(
q^{\prime }\right) \rangle  &=&\left[ \frac{4}{3}\Omega ^{2}\left\vert G_{%
\text{sound}}(q)\right\vert ^{2}\mathfrak{\hat{q}}^{\mu }\mathfrak{\hat{q}}%
^{\nu }-\left\vert G_{\text{shear}}(q)\right\vert ^{2}\Delta _{\bot }^{\mu
\nu }\right] \frac{2\eta T\mathfrak{q}^{2}}{\left( \varepsilon +P\right) ^{2}%
}(2\pi )^{4}\delta ^{(4)}(q+q^{\prime }). 
\end{eqnarray}
Above, we defined the normalized wave-vector, $\mathfrak{\hat{q}%
}^{\mu }=\mathfrak{q}^{\mu }/\mathfrak{q}$, and introduced the following retarded Green's functions, 
\begin{equation}
G_{\text{sound}}(q)=\frac{1}{\left( 1+i\tau _{\pi }\Omega \right) \left(
\Omega ^{2}-\frac{\mathfrak{q}^{2}}{3}\right) -\frac{4}{3}i\tau _{\eta
}\Omega \mathfrak{q}^{2}},\qquad G_{\text{shear}}(q)=\frac{1}{\left( 1+i\tau
_{\pi }\Omega \right) \Omega -i\tau _{\eta }\mathfrak{q}^{2}}.
\label{Gsoundshear}
\end{equation}%
We have checked that when $\tau_\pi\to 0$, our symmetrized correlation functions reduce to the standard results obtained using Navier-Stokes theory \cite{Kovtun:2012rj}.
 
In the linear regime, only the shear-stress tensor correlator contributes, and we find that
\begin{equation}
\tilde{G}\left( q\right) =\frac{4\eta T}{5} \left( \Omega ^{2}-\frac{%
\mathfrak{q}^{2}}{3}\right) ^{2}\left\vert G_{\text{sound}}(q)\right\vert
^{2}+\frac{8\eta T}{5}\Omega ^{2}\left\vert G_{\text{shear}}(q)\right\vert ^{2}+\frac{2}{
1+\tau _{\pi }^{2}\Omega ^{2}} .
\label{Gpipi}
\end{equation}
At vanishing wavenumbers, $\mathfrak{q}=0$, the correlation function \eqref{Gpipi} simplifies to,
\begin{equation}
\tilde{G}(\Omega,0) =\frac{4\eta T}{1+\tau _{\pi }^{2}\Omega ^{2}}.
\label{Gq}
\end{equation}
We note that $\tilde{G}(\Omega,0)$ is directly related to the spectral density, $\mathcal{G}(\Omega)$, calculated in the previous section via metric perturbations, see Eq.\ \eqref{IS_GreenFunction},
\begin{equation}
\tilde{G}\left( q\right) =\frac{2T}{\Omega }\mathcal{G}\left( q\right) .
\label{fluc_diss}
\end{equation}
This is nothing but the fluctuation-dissipation relation \cite{Callen:1951vq,Kubo:1957mj}, and its recovery serves as a consistency check. 

We now use the fluctuation-dissipation relation to re-express our definition of the transport coefficients as,
\begin{equation}
4\eta T \equiv \lim_{\Omega \rightarrow 0}\tilde{G}\left( \Omega ,0\right), \,\,\,\,
\frac{1}{\tau _{\pi }}\equiv\frac{1}{2\eta T }\int \frac{d\Omega }{2\pi }\tilde{G}\left( \Omega ,0\right).
\label{prescription0}
\end{equation}
We note that the ratio $\eta/\tau_\pi$ above is determined by the equal-time symmetric correlator of the shear-stress tensor. The advantage of calculating transport coefficients using the symmetrized correlator, as in \cite{Kovtun:2011np}, is that incorporating the leading order effects of hydrodynamic fluctuations becomes straightforward. This is done by simply considering the nonlinear contributions to the correlation function from the 4-velocity correlators. In this case, the correlator at vanishing $\mathfrak{q}$ is given by the expression (computed in the local rest frame of the fluid),
\begin{eqnarray}
\tilde{G}\left( \Omega ,0\right) &=&\frac{4\eta T}{1+\tau _{\pi }^{2}\Omega
^{2}}
+
\frac{256\left( \eta T\right) ^{2}}{135\left( \varepsilon +P\right) ^{2}}\int \frac{d\omega d^{3}\mathfrak{p}}{(2\pi )^{4}}%
\mathfrak{p}^{4}\omega ^{2}\left\vert G_{\text{sound}%
}(\omega ,\mathfrak{p})\right\vert ^{2}\left( \omega -\Omega \right)
^{2}\left\vert G_{\text{sound}}(\omega -\Omega ,\mathfrak{p})\right\vert
^{2}  \notag
\\
&+&
\frac{56\left( \eta T\right)^{2}}{15\left( \varepsilon +P\right) ^{2}}\int \frac{d\omega d^{3}\mathfrak{p}}{(2\pi )^{4}}%
\mathfrak{p}^{4} \left\vert G_{\text{shear}}(\omega ,\mathfrak{p}%
)\right\vert ^{2}\left\vert G_{\text{shear}}(\omega -\Omega ,\mathfrak{p}%
)\right\vert^{2}  \notag
\\
&+&\frac{64\left( \eta T\right)^{2}}{15\left( \varepsilon +P\right) ^{2}} \int \frac{d\omega d^{3}\mathfrak{p}}{(2\pi )^{4}%
}\mathfrak{p}^{4} \omega ^{2}\left\vert G_{\text{sound}}(\omega ,%
\mathfrak{p})\right\vert ^{2}\left\vert G_{\text{shear}}(\omega -\Omega ,%
\mathfrak{p})\right\vert^{2} ,  \label{correlatorzao}
\end{eqnarray}
where we used the leading order expression for the 4-velocity correlator,
\begin{eqnarray}
\langle \delta \tilde{u}^{\mu }\left( q\right) \delta \tilde{u}^{\nu }\left(
q^{\prime }\right) \rangle  &=&\left[ \frac{4}{3}\Omega ^{2}\left\vert G_{%
\text{sound}}(q)\right\vert ^{2}\mathfrak{\hat{q}}^{\mu }\mathfrak{\hat{q}}%
^{\nu }-\left\vert G_{\text{shear}}(q)\right\vert ^{2}\Delta _{\bot }^{\mu
\nu }\right] \frac{2\eta T\mathfrak{q}^{2}}{\left( \varepsilon +P\right) ^{2}%
}(2\pi )^{4}\delta ^{(4)}(q+q^{\prime }). 
\end{eqnarray}

Corrections to the shear viscosity and relaxation time can be obtained by replacing \eqref{correlatorzao} in \eqref{prescription0}. Performing the frequency integrals exactly using the residue theorem and the momentum integrals with a hard UV cutoff $\Lambda$, one finds the following leading-order expression for the shear viscosity,
\begin{equation}
\eta_{\mathrm{LO}} = \eta + \frac{17 T \Lambda (\varepsilon+P) }{120\pi^2 \eta}, \label{etaR}
\end{equation}
which reproduces\footnote{In these calculations, one assumes $\Lambda \eta/(\varepsilon+P)\ll 1$, as in \cite{Kovtun:2011np}.} the renormalization of the shear viscosity coefficient due to thermal fluctuations found in \cite{Kovtun:2011np}. A non-analytical contribution to $\tilde{G}(\Omega)$ is found at the next order in the expansion, recovering the well-known long-time tail \cite{Kovtun:2003vj} term given by 
\begin{equation}
\tilde{G}(\Omega) \approx 4 T \eta_{\mathrm{LO}} -\left[7+\left(\frac{3}{2}\right)^{3/2}\right]\frac{T^2 (\varepsilon+P)^{3/2}}{60 \pi \eta^{3/2}}\Omega^{1/2},
\end{equation}
also found in \cite{Kovtun:2011np}. This result illustrates the breakdown of the gradient expansion due to the backreaction of thermal fluctuations. In fact, one can show that the corresponding non-analytical contribution to the full $\tilde{G}^{xyxy}_R$ leads to an infrared divergent contribution \cite{Kovtun:2011np} to the microscopic expression of the relaxation time computed via the Kubo formula proposed in \cite{Baier:2007ix}.  

These issues will not appear when using the definition of relaxation time proposed in Eq.~\eqref{prescription0}. In this case, thermal fluctuations change $\eta/\tau_\pi$ as follows:
\begin{equation}
\frac{\eta}{\tau_\pi}\Big|_{\mathrm{LO}} = \frac{\eta}{\tau_\pi} + \frac{T\Lambda^3}{6\pi^2}.
\end{equation}
Given \eqref{etaR}, the equation above implies that $\tau_\pi$ as defined in our work is also renormalized by stochastic fluctuations. 

We now verify if the renormalization of the transport coefficients spoils the causality condition \eqref{causality}, satisfied in the absence of fluctuations. For this purpose, we calculate the effective energy density of the system in the presence of fluctuations,
defined from $u_{\mu }u_{\nu }\left\langle T^{\mu \nu }\right\rangle =%
\mathcal{E}_{\mathrm{LO}}$. Beyond leading order, the result is
\begin{equation}
\mathcal{E}_{\mathrm{LO}}=\varepsilon +\frac{4\eta T}{\varepsilon +P}\int \frac{d\omega
d^{3}\mathfrak{p}}{\left( 2\pi \right) ^{4}}\mathfrak{p}^{2}\left[ \frac{2}{3%
}\omega ^{2}\left\vert G_{\text{sound}}(\omega ,\mathfrak{p})\right\vert
^{2}+\left\vert G_{\text{shear}}(\omega ,\mathfrak{p})\right\vert
^{2}\right] = \varepsilon + \frac{T\Lambda^3}{2\pi^2}.
\end{equation}
We can now determine
\begin{equation}
\frac{\eta}{\tau_\pi(\varepsilon+P)}\Big|_{\mathrm{LO}} = \frac{\eta}{\tau_\pi(\varepsilon+P)}\left[1+\frac{T\Lambda^3}{6\pi^2(\varepsilon+P)}\frac{1}{\left(\frac{\eta}{\tau_\pi(\varepsilon+P)}\right)}\right]\left[1+\frac{T\Lambda^3}{2\pi^2(\varepsilon+P)} \right]^{-1}.
\end{equation}
One can see that $\frac{\eta}{\tau_\pi(\varepsilon+P)}\Big|_{\mathrm{LO}}$ never exceeds 1/2, if the corresponding tree-level result also respects this bound. This suggests that the causality and stability properties of transient relativistic fluid dynamic theories are preserved once the leading thermal fluctuation effects are considered. It would be interesting to extend the results of this work to consider other sources of dissipation, such as bulk viscosity and diffusion \cite{Abbasi:2022aao}. 

\section{Conclusions}

Previous methods for determining the relaxation time from microscopic theories displayed intrinsic shortcomings with and without the effect of thermal fluctuations. In this paper, we solve this problem by proposing another prescription for computing the relaxation time: we demonstrated that $\eta/\tau_\pi$ can be defined as a thermodynamic quantity given by the equal-time correlator of the shear-stress tensor computed in linear response. This allowed us to obtain an exact expression for $\eta/\tau_\pi$ in kinetic theory that is universal, i.e., it is independent of the type of interactions. Similarly, an exact and universal expression for the relaxation time was obtained for holographic gauge theories using the so-called shear-spectral sum rule derived in Ref.~\cite{Romatschke:2009ng}. In both cases, the resulting relaxation times lead to causal and stable theories of transient hydrodynamics. 

We further investigated our new definition of the relaxation time in stochastic transient hydrodynamics. In contrast to previous methods, we demonstrated that our prescription for $\eta/\tau_\pi$ remains meaningful even when considering the backreaction of thermal fluctuations. As a matter of fact, we calculated the leading corrections to $\eta$ and $\tau_\pi$ for a stochastic transient theory of relativistic hydrodynamics. We showed that thermal fluctuations preserve the causality and stability lower bound on $\eta/\tau_\pi$ in our prescription. 
This suggests that stochastic transient relativistic hydrodynamics is a renormalizable theory.

\section*{Acknowledgements}

We thank P.~Kovtun for discussions. G.S.D. acknowledges support from CNPq and Funda\c c\~ao Carlos Chagas Filho de Amparo \`a Pesquisa do Estado do Rio de Janeiro (FAPERJ), grant No. E-26/202.747/2018.
J.N. was partially supported by the U.S. Department of Energy, Office of
Science, Office for Nuclear Physics under Award No. DE-SC0023861. The authors thank KITP Santa Barbara for its hospitality during ``The Many Faces of Relativistic Fluid Dynamics" Program. This research was partly supported by the National Science Foundation under Grant No. NSF PHY-1748958.

\newpage 

\section{Appendix}

We consider a relativistic dilute gas in global equilibrium with an inverse
temperature $\beta =1/T$, a thermal potential $\alpha =\mu /T$, and a
4-velocity $u^{\mu }$. This system is then driven out of equilibrium by
metric perturbations, $h_{\mu \nu }$, 
\begin{equation}
g_{\mu \nu }=\eta _{\mu \nu }+h_{\mu \nu },
\end{equation}%
with $\eta _{\mu \nu }$ being the Minkowski metric. For the sake of
convenience, we consider homogeneous metric perturbations, with only the
following non-vanishing components in the local rest frame of the system, $%
h_{xy}\left( t\right) =h_{yx}\left( t\right) \neq 0$. In the laboratory
frame, such metric perturbations can be shown to be orthogonal to $u^{\mu }$%
, $u^{\mu }h_{\mu \nu }=0$, and traceless, $\eta _{\mu \nu }h^{\mu \nu }=0$.
This type of perturbation will not modify the temperature, chemical
potential and 4-velocity of the system, leading solely to fluctuations of
the shear-stress tensor, $\pi ^{\mu \nu }$.

The dynamics of a relativistic dilute gas is determined by the relativistic
Boltzmann equation -- an integro-differential equation that describes the
spacetime evolution of the local single-particle momentum distribution, $%
f(t,\mathbf{x},\mathbf{k})\equiv f_{\mathbf{k}}$. The Boltzmann equation in
curved space reads \cite{kremer,DEBBASCH20091079,DEBBASCH20091818,Denicol:2014xca,Denicol:2014tha}, 
\begin{equation}
k^{\mu }\partial _{\mu }f_{\mathbf{k}}-k^{\mu }k^{\nu }\Gamma _{\mu \nu }^{i}%
\frac{\partial f_{\mathbf{k}}}{\partial k^{i}}=C\left[ f_{\mathbf{k}}\right],
\label{Boltz}
\end{equation}%
with $C\left[ f_{\mathbf{k}}\right] $ being the collision term, $k^{\mu }$
the particle 4-momenta, and $\Gamma _{\mu \nu }^{\lambda }$ the Christoffel
symbols%
\begin{equation}
\Gamma _{\mu \nu }^{\lambda }=\frac{1}{2}g^{\lambda \alpha }\left( \partial
_{\mu }g_{\alpha \nu }+\partial _{\nu }g_{\alpha \mu }-\partial _{\alpha
}g_{\mu \nu }\right) .  \label{Christofell}
\end{equation}

The metric perturbations will naturally modify the single-particle
distribution function, which we write in the following general form, 
\begin{equation}
f_{\mathbf{k}}\equiv f_{0\mathbf{k}}\left( 1+\tilde{f}_{0\mathbf{k}}\phi _{%
\mathbf{k}}\right) ,  \label{fk}
\end{equation}%
with $f_{0\mathbf{k}}$ being the global equilibrium distribution, 
\begin{equation}
f_{0\mathbf{k}}\equiv \frac{1}{\exp \left( \beta g_{\mu \nu }u^{\nu }k^{\mu
}-\alpha \right) -a},  \label{f_0k}
\end{equation}%
and $a=+$($-$) for bosons (fermions) or $a=0$ when quantum statistics is
disregarded. We also defined the quantity $\tilde{f}_{0\mathbf{k}}\equiv
1+af_{0\mathbf{k}}$. The function $\phi _{\mathbf{k}}$ quantifies the
deviations from global equilibrium and will be calculated in the remainder
of this section in the linear regime.

The linearized Boltzmann equation is obtained by disregarding any
contribution that is quadratic or of a higher power in metric fluctuations.
The result is, 
\begin{equation}
\hat{k}^{\mu }\partial _{\mu }\phi _{\mathbf{k}}-\beta \hat{k}^{\mu
}\partial _{\mu }\left( u_{0}\delta k^{0}\right) +\beta u_{\lambda }\hat{k}%
^{\mu }\hat{k}^{\nu }\delta \Gamma _{\mu \nu }^{i}\frac{\partial \hat{k}%
^{\lambda }}{\partial k^{i}}=-\hat{L}\phi _{\mathbf{k}},  \label{LinearBoltz}
\end{equation}%
with $\hat{L}$ being the linearized collision operator (in flat space) and $%
\delta \Gamma _{\mu \nu }^{\lambda }$ being the linearized Christoffel
symbol, 
\begin{equation}
\delta \Gamma _{\mu \nu }^{\lambda }\equiv \frac{1}{2}\eta ^{\lambda \alpha
}\left( \partial _{\mu }h_{\alpha \nu }+\partial _{\nu }h_{\alpha \mu
}-\partial _{\alpha }h_{\mu \nu }\right) .  \label{Linear_Christ}
\end{equation}%
We further defined the fluctuation of the particle's energy up to first
order in metric perturbations, as 
\begin{equation}
\delta k^{0}\equiv k^{0}-\hat{k}^{0}=-\frac{1}{2\hat{k}^{0}}\hat{k}^{\alpha }%
\hat{k}^{\beta }h_{\alpha \beta },  \label{delta_k0}
\end{equation}%
where we introduced the following notation for the 4-momentum in flat space, 
\begin{equation}
\hat{k}^{\mu }\equiv \left( \sqrt{\left\vert \mathbf{k}\right\vert ^{2}+m^{2}%
},\mathbf{k}\right) .  \label{flat_4k}
\end{equation}%
Finally, we remark that $\hat{L}$\ is a linear operator that vanishes when
applied to quantities that are conserved in microscopic collisions, e.g. $%
\hat{L}\hat{k}^{\mu }=0$.

Using Eqs. \eqref{Linear_Christ} and \eqref{delta_k0}, we can simplify the
linearized Boltzmann equation and recast it in the following form: %
\begin{equation}
\hat{k}^{\lambda }\partial _{\lambda }\phi _{\mathbf{k}}-\frac{\beta }{2}%
\hat{k}^{\left\langle \alpha \right. }\hat{k}^{\left. \beta \right\rangle
}u^{\lambda }\partial _{\lambda }h_{\alpha \beta }=-\hat{L}\phi _{\mathbf{k}%
},  \label{Linear_eq}
\end{equation}%
where we also used the following result, 
\begin{equation}
u_{\lambda }\frac{\partial k^{\lambda }}{\partial k^{i}}=-\frac{u_{0}}{k_{0}}%
k_{i}+u_{i}.
\end{equation}%
To write \eqref{Linear_eq} in this form, we used that the metric
perturbations are orthogonal to $u^{\mu }$ and traceless. In Fourier space,
Eq. \eqref{Linear_eq} becomes
\begin{equation}
\left( iq_{\lambda }\hat{k}^{\lambda }+\hat{L}\right) \tilde{\phi}_{\mathbf{k%
}}=\frac{i\beta }{2}\Omega \hat{k}^{\left\langle \alpha \right. }\hat{k}%
^{\left. \beta \right\rangle }\tilde{h}_{\alpha \beta },
\end{equation}%
with $\Omega \equiv u_{\lambda }\hat{k}^{\lambda }$ being the frequency in
the local rest frame of the system. We note that terms $\sim k^{\left\langle \mu \right. }k^{\left. \nu \right\rangle }$ are not zero modes of $\hat{L}$ and, thus, the operator $iq_{\lambda }\hat{k}^{\lambda }+\hat{L}$ can be inverted in this subspace. The particular solution for $\tilde{\phi}%
_{\mathbf{k}}$ can then be cast in the following simple form,
\begin{equation}
\tilde{\phi}_{\mathbf{k}}=\frac{\beta }{2}\frac{i\Omega }{iq_{\lambda }\hat{k%
}^{\lambda }+\hat{L}}\hat{k}^{\left\langle \alpha \right. }\hat{k}^{\left.
\beta \right\rangle }\tilde{h}_{\alpha \beta }.  \label{Particular_Sol}
\end{equation}

The kinetic expression for the energy-momentum tensor is, \begin{eqnarray}
T^{\mu \nu } =\int dK k^{\mu }k^{\nu }f_{\mathbf{k}},
\end{eqnarray}%
where $dK\equiv d^3k/[(2\pi)^3k^0]$. Using the solution for the single-particle distribution found in the previous section, the energy-momentum tensor fluctuations can be cast in the following form, 
\begin{equation}
\delta \tilde{T}^{\mu \nu }=\frac{1}{2}\tilde{G}_{R}^{\mu \nu \alpha \beta
}\tilde{h}_{\alpha \beta },  \label{Linear_Response}
\end{equation}%
and we identify the retarded Green's function as%
\begin{equation}
\tilde{G}_{R}^{\mu \nu \alpha \beta }=\beta \int dKf_{0\mathbf{k}}\tilde{f}%
_{0\mathbf{k}}\hat{k}^{\mu }\hat{k}^{\nu }\frac{i\Omega }{iq_{\lambda }\hat{k%
}^{\lambda }+\hat{L}}\hat{k}^{\left\langle \alpha \right. }\hat{k}^{\left.
\beta \right\rangle }+\text{contact terms},
\label{GreenFunc}
\end{equation}%
where we assume that $q_\mu \in \mathbb{R}$. For the sake of convenience, we define the imaginary part of this Green's
function,
\begin{equation}
\mathcal{G}^{\mu \nu \alpha \beta }\equiv \,\textrm{Im}\,\tilde{G}_{R}^{\mu \nu
\alpha \beta }=\beta \, \textrm{Im}\int dKf_{0\mathbf{k}}\tilde{f}_{0\mathbf{k}}%
\hat{k}^{\mu }\hat{k}^{\nu }\frac{i\Omega }{iq_{\lambda }\hat{k}^{\lambda }+%
\hat{L}}\hat{k}^{\left\langle \alpha \right. }\hat{k}^{\left. \beta
\right\rangle }.
\end{equation}
We note that the contact terms mentioned in Eq.~\eqref{GreenFunc} have no imaginary part and do not contribute to $\mathcal{G}^{\mu \nu \alpha \beta }$.

We now decompose the Fourier 4-momentum, $q^{\mu }$, in terms of its
components parallel and orthogonal to the 4-velocity, $u^{\mu }$, 
\begin{equation}
q^{\mu }=\Omega u^{\mu }+\mathfrak{q}^{\mu }=\Omega u^{\mu }+\mathfrak{q\hat{%
q}}^{\mu },
\end{equation}%
where $\mathfrak{q}^{\mu }=$ $\Delta ^{\mu \nu }q_{\nu }$, $\mathfrak{\hat{q}%
}^{\mu }=\mathfrak{q}^{\mu }/\mathfrak{q}$, and $\mathfrak{q}_{\mu }%
\mathfrak{q}^{\mu }=-\mathfrak{q}^{2}$. We note
that for a static equilibrium state, $u^{\mu }=(1,0,0,0)$, $\Omega =\omega $
and $\mathfrak{q}^{\mu }=(0,\mathbf{q})$. We remark that all the
following calculations will be performed in the limit of vanishing
wavenumber, $\mathfrak{q}\rightarrow 0$. In this case, the imaginary part of
the retarded Green's function can be expressed in the straightforward form,
\begin{equation}
\mathcal{G}^{\mu \nu \alpha \beta }\left( \Omega,0 \right) =\beta \,\text{Im}%
\int dKf_{0\mathbf{k}}\tilde{f}_{0\mathbf{k}}\hat{k}^{\left\langle \mu
\right. }\hat{k}^{\left. \nu \right\rangle }\frac{i\Omega }{i\Omega E_{%
\mathbf{k}}+\hat{L}}\hat{k}^{\left\langle \alpha \right. }\hat{k}^{\left.
\beta \right\rangle }\equiv \mathcal{G}\left( \Omega \right) \Delta ^{\mu
\nu \alpha \beta },  \label{Nice_Relation}
\end{equation}%
where we defined the energy in the local rest frame of the system, $E_{%
\mathbf{k}}\equiv u_{\mu }k^{\mu }$, and introduced the Lorentz scalar
function, 
\begin{eqnarray}
\mathcal{G}\left( \Omega \right) =\frac{\beta }{5}\,\text{Im}\int dKf_{0%
\mathbf{k}}\tilde{f}_{0\mathbf{k}}k_{\left\langle \mu \right. }k_{\left. \nu
\right\rangle }\frac{i\Omega }{i\Omega E_{\mathbf{k}}+\hat{L}}%
k^{\left\langle \mu \right. }k^{\left. \nu \right\rangle }  \label{S_Q}.
\end{eqnarray}
The integral over momentum defined above exists and, thus, can be, in principle, exchanged with the operation of taking the imaginary part. Before doing so, we rewrite the expression for $\mathcal{G}$ factorizing $E_{\mathbf{k}}$ from the denominator and defining the operator $\hat{D} \equiv E_{\mathbf{k}}^{-1}\hat{L}$, 
\begin{eqnarray}
\mathcal{G}\left( \Omega \right) =\frac{\beta }{5}\,\text{Im}\int dKf_{0%
\mathbf{k}}\tilde{f}_{0\mathbf{k}}k_{\left\langle \mu \right. }k_{\left. \nu
\right\rangle }\frac{i\Omega }{i\Omega+\hat{D}}%
E_{\mathbf{k}}^{-1}k^{\left\langle \mu \right. }k^{\left. \nu \right\rangle } .
\end{eqnarray}
We then take the imaginary part of the integrand\footnote{We use the traditional technique of multiplying the numerator and denominator of the integrand by $-i\Omega + \hat{D}$ so that the denominator is real. This is possible since $\hat{D}$ and $\Omega$ commute.}, to find
\begin{eqnarray}
\mathcal{G}\left( \Omega \right) =\frac{\beta }{5}\int dKf_{0%
\mathbf{k}}\tilde{f}_{0\mathbf{k}}k_{\left\langle \mu \right. }k_{\left. \nu
\right\rangle }\frac{\Omega }{\Omega^2+\hat{D}^2}%
\hat{D}(E_{\mathbf{k}}^{-1}k^{\left\langle \mu \right. }k^{\left. \nu \right\rangle }).  \label{S_Q2}
\end{eqnarray}

The transport properties of a fluid related to the shear-stress tensor fluctuations
emerge from $\mathcal{G}$. For instance, the shear viscosity coefficient, $\eta $, is determined from the
vanishing frequency limit of the ratio $\mathcal{G}\left( \Omega \right)
/\Omega $,

\begin{equation}
2\eta =\lim_{\Omega \rightarrow 0}\frac{\mathcal{G}\left( \Omega \right) }{%
\Omega }=\frac{\beta }{5}\int \frac{d^{3}k}{\left( 2\pi \right) ^{3}\hat{k}%
^{0}}f_{0\mathbf{k}}\tilde{f}_{0\mathbf{k}}k_{\left\langle \mu \right.
}k_{\left. \nu \right\rangle }\frac{1}{\hat{L}}k^{\left\langle \mu \right.
}k^{\left. \nu \right\rangle }.
\end{equation}%
Naturally, an actual value for the shear viscosity can only be obtained once
the operator $\hat{L}$ is specified, describing a particular choice of interaction.
Nevertheless, the calculation of the shear viscosity is often very
complicated since the linear operator $\hat{L}$ must be inverted. This
procedure must be carried out numerically in most cases. We note that for
massless and classical scalar fields self-interacting via a quartic
potential, the spectrum of $\hat{L}$ was determined analytically in Ref.~%
\cite{Denicol:2022bsq}. For instance, it was shown that $k^{\left\langle \mu
\right. }k^{\left. \nu \right\rangle }$ is an eigenfunction of $\hat{L}$,
i.e., $\hat{L}k^{\left\langle \mu \right. }k^{\left. \nu \right\rangle
}=\chi k^{\left\langle \mu \right. }k^{\left. \nu \right\rangle }$. The
eigenvalue $\chi $ was determined exactly, 
\begin{equation}
\chi =\frac{g\exp \alpha }{12\pi ^{2}\beta ^{2}},
\end{equation}%
where we further defined $g=\lambda ^{2}/(32\pi )$, with $\lambda $ being
the coupling constant. For this interaction, the shear viscosity can then be
calculated analytically from the Kubo formula,

\begin{equation}
\eta =\frac{\beta }{10\chi} \int dK \,f_{0\mathbf{k}}k_{\left\langle \mu
\right. }k_{\left. \nu \right\rangle }k^{\left\langle \mu \right. }k^{\left.
\nu \right\rangle }=\frac{48}{g}T^{3}.
\end{equation}%
This result is identical to the shear viscosity calculated in \cite%
{Denicol:2022bsq}, for the same system, which serves as a consistency check.

Next, we calculate the integral over frequency of $\mathcal{G}\left( \Omega
\right) /\Omega $,

\begin{equation}
\int_{-\infty }^{\infty }\frac{d\Omega }{2\pi }\frac{\mathcal{G}%
\left( \Omega \right) }{\Omega }=\frac{\beta }{5}\int dKf_{0\mathbf{k}}%
\tilde{f}_{0\mathbf{k}}k_{\left\langle \mu \right. }k_{\left. \nu
\right\rangle }\int_{-\infty }^{\infty }\frac{d\Omega }{2\pi }\frac{1}{%
1+\left( \Omega \hat{L}^{-1}E_{\mathbf{k}}\right)^{2}}\hat{L}%
^{-1}k^{\left\langle \mu \right. }k^{\left. \nu \right\rangle }.
\end{equation}%
The integral over frequency can be calculated analytically using the
relation, 
\begin{equation}
\int \frac{d\Omega }{\pi }\frac{1}{1+(\Omega x)^2}=\frac{1}{|x|}.
\end{equation}%
We then obtain the final result, 
\begin{equation}
\int_{-\infty }^{\infty }\frac{d\Omega }{2\pi }\frac{\mathcal{G}%
\left( \Omega \right) }{\Omega }=\frac{\beta }{10}\int dKf_{0\mathbf{k}}%
\tilde{f}_{0\mathbf{k}}k_{\left\langle \mu \right. }k_{\left. \nu
\right\rangle }\frac{1}{\hat{L}^{-1}E_{\mathbf{k}}}\hat{L}%
^{-1}k^{\left\langle \mu \right. }k^{\left. \nu \right\rangle }=\frac{\beta 
}{10}\int dKf_{0\mathbf{k}}\tilde{f}_{0\mathbf{k}}E_{\mathbf{k}%
}^{-1}k_{\left\langle \mu\right. }k_{\left. \nu \right\rangle }
k^{\left\langle \mu \right. }k^{\left. \nu \right\rangle }.  \label{Sum_Rule}
\end{equation}%
The remarkable aspect of this expression is its universality: the collision operator exactly disappears from the expression, and this integral of the
retarded Green's function becomes a purely thermodynamic function. In the massless limit, $\chi $ can be
calculated explicitly in terms of the energy density and pressure,%
\begin{equation}
\int_{-\infty }^{\infty }\frac{d\Omega }{2\pi }\frac{\mathcal{G}%
\left( \Omega \right) }{\Omega } =\frac{\varepsilon +P}{5}.
\end{equation}

In the main text, we argued that transient fluid dynamic theories satisfy this relation exactly, leading to a redefinition of the shear relaxation time. In this case, we obtained the following expression for $\eta / \tau_\pi$, 
\begin{equation}
\frac{1}{\tau _{\pi }}=\frac{1}{\eta }\int_{-\infty }^{\infty }\frac{d\Omega 
}{2\pi }\frac{\mathcal{G}\left( \Omega ,0\right) }{\Omega },
\label{relax_time}
\end{equation}%
with the shear viscosity being determined by the usual Kubo formula. In kinetic theory, we then obtain the universal expression for the ratio $\eta / \tau_\pi$,
\begin{equation}
\frac{\eta}{\tau _{\pi }}=\frac{\beta 
}{10}\int dKf_{0\mathbf{k}}\tilde{f}_{0\mathbf{k}}E_{\mathbf{k}%
}^{-1}k_{\left\langle \mu\right. }k_{\left. \nu \right\rangle }
k^{\left\langle \mu \right. }k^{\left. \nu \right\rangle }.
\label{relax_time_kinetic}
\end{equation}
In the massless limit, this leads to the following microscopic expression for the relaxation time,
\begin{equation}
\tau _{\pi }=\frac{5\eta 
}{\varepsilon + P}.
\label{relax_time_kinetic_massless}
\end{equation}

\bibliography{References.bib}

\def\cprime{$'$}
\begin{thebibliography}{63}%
\makeatletter
\providecommand \@ifxundefined [1]{%
 \@ifx{#1\undefined}
}%
\providecommand \@ifnum [1]{%
 \ifnum #1\expandafter \@firstoftwo
 \else \expandafter \@secondoftwo
 \fi
}%
\providecommand \@ifx [1]{%
 \ifx #1\expandafter \@firstoftwo
 \else \expandafter \@secondoftwo
 \fi
}%
\providecommand \natexlab [1]{#1}%
\providecommand \enquote  [1]{``#1''}%
\providecommand \bibnamefont  [1]{#1}%
\providecommand \bibfnamefont [1]{#1}%
\providecommand \citenamefont [1]{#1}%
\providecommand \href@noop [0]{\@secondoftwo}%
\providecommand \href [0]{\begingroup \@sanitize@url \@href}%
\providecommand \@href[1]{\@@startlink{#1}\@@href}%
\providecommand \@@href[1]{\endgroup#1\@@endlink}%
\providecommand \@sanitize@url [0]{\catcode `\\12\catcode `\$12\catcode
  `\&12\catcode `\#12\catcode `\^12\catcode `\_12\catcode `\%12\relax}%
\providecommand \@@startlink[1]{}%
\providecommand \@@endlink[0]{}%
\providecommand \url  [0]{\begingroup\@sanitize@url \@url }%
\providecommand \@url [1]{\endgroup\@href {#1}{\urlprefix }}%
\providecommand \urlprefix  [0]{URL }%
\providecommand \Eprint [0]{\href }%
\providecommand \doibase [0]{https://doi.org/}%
\providecommand \selectlanguage [0]{\@gobble}%
\providecommand \bibinfo  [0]{\@secondoftwo}%
\providecommand \bibfield  [0]{\@secondoftwo}%
\providecommand \translation [1]{[#1]}%
\providecommand \BibitemOpen [0]{}%
\providecommand \bibitemStop [0]{}%
\providecommand \bibitemNoStop [0]{.\EOS\space}%
\providecommand \EOS [0]{\spacefactor3000\relax}%
\providecommand \BibitemShut  [1]{\csname bibitem#1\endcsname}%
\let\auto@bib@innerbib\@empty
\bibitem [{\citenamefont {Gale}\ \emph {et~al.}(2013)\citenamefont {Gale},
  \citenamefont {Jeon},\ and\ \citenamefont {Schenke}}]{Gale:2013da}%
  \BibitemOpen
  \bibfield  {author} {\bibinfo {author} {\bibfnamefont {G.}~\bibnamefont
  {Gale}}, \bibinfo {author} {\bibfnamefont {S.}~\bibnamefont {Jeon}},\ and\
  \bibinfo {author} {\bibfnamefont {B.}~\bibnamefont {Schenke}},\ }\bibfield
  {title} {\bibinfo {title} {Hydrodynamic modeling of heavy-ion collisions},\
  }\href {https://doi.org/10.1142/S0217751X13400113} {\bibfield  {journal}
  {\bibinfo  {journal} {Int.\ J.\ Mod.\ Phys.\ A}\ }\textbf {\bibinfo {volume}
  {28}},\ \bibinfo {pages} {1340011} (\bibinfo {year} {2013})},\ \Eprint
  {https://arxiv.org/abs/1301.5893} {arXiv:1301.5893 [nucl-th]} \BibitemShut
  {NoStop}%
\bibitem [{\citenamefont {Heinz}\ and\ \citenamefont
  {Snellings}(2013)}]{Heinz:2013th}%
  \BibitemOpen
  \bibfield  {author} {\bibinfo {author} {\bibfnamefont {U.}~\bibnamefont
  {Heinz}}\ and\ \bibinfo {author} {\bibfnamefont {R.}~\bibnamefont
  {Snellings}},\ }\bibfield  {title} {\bibinfo {title} {Collective flow and
  viscosity in relativistic heavy-ion collisions},\ }\href
  {https://doi.org/10.1146/annurev-nucl-102212-170540} {\bibfield  {journal}
  {\bibinfo  {journal} {Ann. Rev. Nucl. Part. Sci.}\ }\textbf {\bibinfo
  {volume} {63}},\ \bibinfo {pages} {123} (\bibinfo {year} {2013})},\ \Eprint
  {https://arxiv.org/abs/1301.2826} {arXiv:1301.2826 [nucl-th]} \BibitemShut
  {NoStop}%
\bibitem [{\citenamefont {Romatschke}\ and\ \citenamefont
  {Romatschke}(2019)}]{Romatschke:2017ejr}%
  \BibitemOpen
  \bibfield  {author} {\bibinfo {author} {\bibfnamefont {P.}~\bibnamefont
  {Romatschke}}\ and\ \bibinfo {author} {\bibfnamefont {U.}~\bibnamefont
  {Romatschke}},\ }\href@noop {} {\emph {\bibinfo {title} {Relativistic Fluid
  Dynamics In and Out of Equilibrium}}},\ Cambridge Monographs on Mathematical
  Physics\ (\bibinfo  {publisher} {Cambridge University Press},\ \bibinfo
  {year} {2019})\ \Eprint {https://arxiv.org/abs/1712.05815} {arXiv:1712.05815
  [nucl-th]} \BibitemShut {NoStop}%
\bibitem [{\citenamefont {Baiotti}\ and\ \citenamefont
  {Rezzolla}(2017)}]{Baiotti:2016qnr}%
  \BibitemOpen
  \bibfield  {author} {\bibinfo {author} {\bibfnamefont {L.}~\bibnamefont
  {Baiotti}}\ and\ \bibinfo {author} {\bibfnamefont {L.}~\bibnamefont
  {Rezzolla}},\ }\bibfield  {title} {\bibinfo {title} {Binary neutron star
  mergers: a review of einstein's richest laboratory},\ }\href
  {https://doi.org/10.1088/1361-6633/aa67bb} {\bibfield  {journal} {\bibinfo
  {journal} {Rept. Prog. Phys.}\ }\textbf {\bibinfo {volume} {80}},\ \bibinfo
  {pages} {096901} (\bibinfo {year} {2017})},\ \Eprint
  {https://arxiv.org/abs/1607.03540} {arXiv:1607.03540 [gr-qc]} \BibitemShut
  {NoStop}%
\bibitem [{\citenamefont {Rocha}\ \emph
  {et~al.}(2024{\natexlab{a}})\citenamefont {Rocha}, \citenamefont {Wagner},
  \citenamefont {Denicol}, \citenamefont {Noronha},\ and\ \citenamefont
  {Rischke}}]{Rocha:2023ilf}%
  \BibitemOpen
  \bibfield  {author} {\bibinfo {author} {\bibfnamefont {G.~S.}\ \bibnamefont
  {Rocha}}, \bibinfo {author} {\bibfnamefont {D.}~\bibnamefont {Wagner}},
  \bibinfo {author} {\bibfnamefont {G.~S.}\ \bibnamefont {Denicol}}, \bibinfo
  {author} {\bibfnamefont {J.}~\bibnamefont {Noronha}},\ and\ \bibinfo {author}
  {\bibfnamefont {D.~H.}\ \bibnamefont {Rischke}},\ }\bibfield  {title}
  {\bibinfo {title} {{Theories of Relativistic Dissipative Fluid Dynamics}},\
  }\href {https://doi.org/10.3390/e26030189} {\bibfield  {journal} {\bibinfo
  {journal} {Entropy}\ }\textbf {\bibinfo {volume} {26}},\ \bibinfo {pages}
  {189} (\bibinfo {year} {2024}{\natexlab{a}})},\ \Eprint
  {https://arxiv.org/abs/2311.15063} {arXiv:2311.15063 [nucl-th]} \BibitemShut
  {NoStop}%
\bibitem [{\citenamefont {Landau}\ and\ \citenamefont
  {Lifshitz}(1987)}]{LandauLifshitzFluids}%
  \BibitemOpen
  \bibfield  {author} {\bibinfo {author} {\bibfnamefont {L.~D.}\ \bibnamefont
  {Landau}}\ and\ \bibinfo {author} {\bibfnamefont {E.~M.}\ \bibnamefont
  {Lifshitz}},\ }\href@noop {} {\emph {\bibinfo {title} {Fluid Mechanics -
  Volume 6 (Course of Theoretical Physics)}}},\ \bibinfo {edition} {2nd}\ ed.\
  (\bibinfo  {publisher} {Butterworth-Heinemann},\ \bibinfo {year} {1987})\ p.\
  \bibinfo {pages} {552}\BibitemShut {NoStop}%
\bibitem [{\citenamefont {Eckart}(1940)}]{EckartViscous}%
  \BibitemOpen
  \bibfield  {author} {\bibinfo {author} {\bibfnamefont {C.}~\bibnamefont
  {Eckart}},\ }\bibfield  {title} {\bibinfo {title} {The thermodynamics of
  irreversible processes {III}. {R}elativistic theory of the simple fluid},\
  }\href@noop {} {\bibfield  {journal} {\bibinfo  {journal} {Physical Review}\
  }\textbf {\bibinfo {volume} {58}},\ \bibinfo {pages} {919} (\bibinfo {year}
  {1940})}\BibitemShut {NoStop}%
\bibitem [{\citenamefont {Pichon}(1965)}]{PichonViscous}%
  \BibitemOpen
  \bibfield  {author} {\bibinfo {author} {\bibfnamefont {G.}~\bibnamefont
  {Pichon}},\ }\bibfield  {title} {\bibinfo {title} {{\'E}tude relativiste de
  fluides visqueux et charg{\'e}s},\ }\href
  {http://www.numdam.org/item/AIHPA_1965__2_1_21_0} {\bibfield  {journal}
  {\bibinfo  {journal} {Annales de l'I.H.P. Physique th{\'e}orique}\ }\textbf
  {\bibinfo {volume} {2}},\ \bibinfo {pages} {21} (\bibinfo {year}
  {1965})}\BibitemShut {NoStop}%
\bibitem [{\citenamefont {Hiscock}\ and\ \citenamefont
  {Lindblom}(1985)}]{Hiscock_Lindblom_instability_1985}%
  \BibitemOpen
  \bibfield  {author} {\bibinfo {author} {\bibfnamefont {W.~A.}\ \bibnamefont
  {Hiscock}}\ and\ \bibinfo {author} {\bibfnamefont {L.}~\bibnamefont
  {Lindblom}},\ }\bibfield  {title} {\bibinfo {title} {Generic instabilities in
  first-order dissipative fluid theories},\ }\href@noop {} {\bibfield
  {journal} {\bibinfo  {journal} {Phys. Rev. D}\ }\textbf {\bibinfo {volume}
  {31}},\ \bibinfo {pages} {725} (\bibinfo {year} {1985})}\BibitemShut
  {NoStop}%
\bibitem [{\citenamefont {Denicol}\ \emph {et~al.}(2008)\citenamefont
  {Denicol}, \citenamefont {Kodama}, \citenamefont {Koide},\ and\ \citenamefont
  {Mota}}]{Denicol:2008ha}%
  \BibitemOpen
  \bibfield  {author} {\bibinfo {author} {\bibfnamefont {G.~S.}\ \bibnamefont
  {Denicol}}, \bibinfo {author} {\bibfnamefont {T.}~\bibnamefont {Kodama}},
  \bibinfo {author} {\bibfnamefont {T.}~\bibnamefont {Koide}},\ and\ \bibinfo
  {author} {\bibfnamefont {P.}~\bibnamefont {Mota}},\ }\bibfield  {title}
  {\bibinfo {title} {Stability and causality in relativistic dissipative
  hydrodynamics},\ }\href {https://doi.org/10.1088/0954-3899/35/11/115102}
  {\bibfield  {journal} {\bibinfo  {journal} {J. Phys.}\ }\textbf {\bibinfo
  {volume} {G35}},\ \bibinfo {pages} {115102} (\bibinfo {year} {2008})},\
  \Eprint {https://arxiv.org/abs/0807.3120} {arXiv:0807.3120 [hep-ph]}
  \BibitemShut {NoStop}%
\bibitem [{\citenamefont {Pu}\ \emph {et~al.}(2010)\citenamefont {Pu},
  \citenamefont {Koide},\ and\ \citenamefont {Rischke}}]{Pu:2009fj}%
  \BibitemOpen
  \bibfield  {author} {\bibinfo {author} {\bibfnamefont {S.}~\bibnamefont
  {Pu}}, \bibinfo {author} {\bibfnamefont {T.}~\bibnamefont {Koide}},\ and\
  \bibinfo {author} {\bibfnamefont {D.~H.}\ \bibnamefont {Rischke}},\
  }\bibfield  {title} {\bibinfo {title} {Does stability of relativistic
  dissipative fluid dynamics imply causality?},\ }\href
  {https://doi.org/10.1103/PhysRevD.81.114039} {\bibfield  {journal} {\bibinfo
  {journal} {Phys. Rev.}\ }\textbf {\bibinfo {volume} {D81}},\ \bibinfo {pages}
  {114039} (\bibinfo {year} {2010})},\ \Eprint
  {https://arxiv.org/abs/0907.3906} {arXiv:0907.3906 [hep-ph]} \BibitemShut
  {NoStop}%
\bibitem [{\citenamefont {Gavassino}(2022)}]{Gavassino:2021owo}%
  \BibitemOpen
  \bibfield  {author} {\bibinfo {author} {\bibfnamefont {L.}~\bibnamefont
  {Gavassino}},\ }\bibfield  {title} {\bibinfo {title} {{Can We Make Sense of
  Dissipation without Causality?}},\ }\href
  {https://doi.org/10.1103/PhysRevX.12.041001} {\bibfield  {journal} {\bibinfo
  {journal} {Phys. Rev. X}\ }\textbf {\bibinfo {volume} {12}},\ \bibinfo
  {pages} {041001} (\bibinfo {year} {2022})},\ \Eprint
  {https://arxiv.org/abs/2111.05254} {arXiv:2111.05254 [gr-qc]} \BibitemShut
  {NoStop}%
\bibitem [{\citenamefont {Israel}(1976{\natexlab{a}})}]{Israel:1976tn}%
  \BibitemOpen
  \bibfield  {author} {\bibinfo {author} {\bibfnamefont {W.}~\bibnamefont
  {Israel}},\ }\bibfield  {title} {\bibinfo {title} {{Nonstationary
  irreversible thermodynamics: A Causal relativistic theory}},\ }\href
  {https://doi.org/10.1016/0003-4916(76)90064-6} {\bibfield  {journal}
  {\bibinfo  {journal} {Annals Phys.}\ }\textbf {\bibinfo {volume} {100}},\
  \bibinfo {pages} {310} (\bibinfo {year} {1976}{\natexlab{a}})}\BibitemShut
  {NoStop}%
\bibitem [{\citenamefont {Israel}\ and\ \citenamefont {Stewart}(1979)}]{MIS-6}%
  \BibitemOpen
  \bibfield  {author} {\bibinfo {author} {\bibfnamefont {W.}~\bibnamefont
  {Israel}}\ and\ \bibinfo {author} {\bibfnamefont {J.~M.}\ \bibnamefont
  {Stewart}},\ }\bibfield  {title} {\bibinfo {title} {Transient relativistic
  thermodynamics and kinetic theory},\ }\href@noop {} {\bibfield  {journal}
  {\bibinfo  {journal} {Ann. Phys.}\ }\textbf {\bibinfo {volume} {118}},\
  \bibinfo {pages} {341} (\bibinfo {year} {1979})}\BibitemShut {NoStop}%
\bibitem [{\citenamefont {Bemfica}\ \emph {et~al.}(2018)\citenamefont
  {Bemfica}, \citenamefont {Disconzi},\ and\ \citenamefont
  {Noronha}}]{Bemfica:2017wps}%
  \BibitemOpen
  \bibfield  {author} {\bibinfo {author} {\bibfnamefont {F.~S.}\ \bibnamefont
  {Bemfica}}, \bibinfo {author} {\bibfnamefont {M.~M.}\ \bibnamefont
  {Disconzi}},\ and\ \bibinfo {author} {\bibfnamefont {J.}~\bibnamefont
  {Noronha}},\ }\bibfield  {title} {\bibinfo {title} {Causality and existence
  of solutions of relativistic viscous fluid dynamics with gravity},\ }\href
  {https://doi.org/10.1103/PhysRevD.98.104064} {\bibfield  {journal} {\bibinfo
  {journal} {Physical Review D}\ }\textbf {\bibinfo {volume} {98}},\ \bibinfo
  {pages} {104064 (26 pages)} (\bibinfo {year} {2018})},\ \Eprint
  {https://arxiv.org/abs/1708.06255} {arXiv:1708.06255 [gr-qc]} \BibitemShut
  {NoStop}%
\bibitem [{\citenamefont {Kovtun}(2019)}]{Kovtun:2019hdm}%
  \BibitemOpen
  \bibfield  {author} {\bibinfo {author} {\bibfnamefont {P.}~\bibnamefont
  {Kovtun}},\ }\bibfield  {title} {\bibinfo {title} {First-order relativistic
  hydrodynamics is stable},\ }\href {https://doi.org/10.1007/JHEP10(2019)034}
  {\bibfield  {journal} {\bibinfo  {journal} {JHEP}\ }\textbf {\bibinfo
  {volume} {10}},\ \bibinfo {pages} {034}},\ \Eprint
  {https://arxiv.org/abs/1907.08191} {arXiv:1907.08191 [hep-th]} \BibitemShut
  {NoStop}%
\bibitem [{\citenamefont {Bemfica}\ \emph {et~al.}(2019)\citenamefont
  {Bemfica}, \citenamefont {Disconzi},\ and\ \citenamefont
  {Noronha}}]{Bemfica:2019knx}%
  \BibitemOpen
  \bibfield  {author} {\bibinfo {author} {\bibfnamefont {F.~S.}\ \bibnamefont
  {Bemfica}}, \bibinfo {author} {\bibfnamefont {M.~M.}\ \bibnamefont
  {Disconzi}},\ and\ \bibinfo {author} {\bibfnamefont {J.}~\bibnamefont
  {Noronha}},\ }\bibfield  {title} {\bibinfo {title} {Nonlinear causality of
  general first-order relativistic viscous hydrodynamics},\ }\href
  {https://doi.org/10.1103/PhysRevD.100.104020} {\bibfield  {journal} {\bibinfo
   {journal} {Physisical Review D}\ }\textbf {\bibinfo {volume} {100}},\
  \bibinfo {pages} {104020 (13 pages)} (\bibinfo {year} {2019})},\ \Eprint
  {https://arxiv.org/abs/1907.12695} {arXiv:1907.12695 [gr-qc]} \BibitemShut
  {NoStop}%
\bibitem [{\citenamefont {Hoult}\ and\ \citenamefont
  {Pavel}(2020)}]{Hoult:2020eho}%
  \BibitemOpen
  \bibfield  {author} {\bibinfo {author} {\bibfnamefont {R.~E.}\ \bibnamefont
  {Hoult}}\ and\ \bibinfo {author} {\bibfnamefont {K.}~\bibnamefont {Pavel}},\
  }\bibfield  {title} {\bibinfo {title} {Stable and causal relativistic
  navier-stokes equations},\ }\href@noop {} {\  (\bibinfo {year} {2020})},\
  \Eprint {https://arxiv.org/abs/2004.04102} {arXiv:2004.04102 [hep-th]}
  \BibitemShut {NoStop}%
\bibitem [{\citenamefont {Bemfica}\ \emph {et~al.}(2022)\citenamefont
  {Bemfica}, \citenamefont {Disconzi},\ and\ \citenamefont
  {Noronha}}]{Bemfica:2020zjp}%
  \BibitemOpen
  \bibfield  {author} {\bibinfo {author} {\bibfnamefont {F.~S.}\ \bibnamefont
  {Bemfica}}, \bibinfo {author} {\bibfnamefont {M.~M.}\ \bibnamefont
  {Disconzi}},\ and\ \bibinfo {author} {\bibfnamefont {J.}~\bibnamefont
  {Noronha}},\ }\bibfield  {title} {\bibinfo {title} {{First-Order
  General-Relativistic Viscous Fluid Dynamics}},\ }\href
  {https://doi.org/10.1103/PhysRevX.12.021044} {\bibfield  {journal} {\bibinfo
  {journal} {Phys. Rev. X}\ }\textbf {\bibinfo {volume} {12}},\ \bibinfo
  {pages} {021044} (\bibinfo {year} {2022})},\ \Eprint
  {https://arxiv.org/abs/2009.11388} {arXiv:2009.11388 [gr-qc]} \BibitemShut
  {NoStop}%
\bibitem [{\citenamefont {Hiscock}\ and\ \citenamefont
  {Lindblom}(1983)}]{Hiscock_Lindblom_stability_1983}%
  \BibitemOpen
  \bibfield  {author} {\bibinfo {author} {\bibfnamefont {W.~A.}\ \bibnamefont
  {Hiscock}}\ and\ \bibinfo {author} {\bibfnamefont {L.}~\bibnamefont
  {Lindblom}},\ }\bibfield  {title} {\bibinfo {title} {Stability and causality
  in dissipative relativistic fluids},\ }\href@noop {} {\bibfield  {journal}
  {\bibinfo  {journal} {Annals of Physics}\ }\textbf {\bibinfo {volume}
  {151}},\ \bibinfo {pages} {466} (\bibinfo {year} {1983})}\BibitemShut
  {NoStop}%
\bibitem [{\citenamefont {Wagner}\ and\ \citenamefont
  {Gavassino}(2024)}]{Wagner:2023jgq}%
  \BibitemOpen
  \bibfield  {author} {\bibinfo {author} {\bibfnamefont {D.}~\bibnamefont
  {Wagner}}\ and\ \bibinfo {author} {\bibfnamefont {L.}~\bibnamefont
  {Gavassino}},\ }\bibfield  {title} {\bibinfo {title} {{Regime of
  applicability of Israel-Stewart hydrodynamics}},\ }\href
  {https://doi.org/10.1103/PhysRevD.109.016019} {\bibfield  {journal} {\bibinfo
   {journal} {Phys. Rev. D}\ }\textbf {\bibinfo {volume} {109}},\ \bibinfo
  {pages} {016019} (\bibinfo {year} {2024})},\ \Eprint
  {https://arxiv.org/abs/2309.14828} {arXiv:2309.14828 [nucl-th]} \BibitemShut
  {NoStop}%
\bibitem [{\citenamefont {Kovtun}(2012)}]{Kovtun:2012rj}%
  \BibitemOpen
  \bibfield  {author} {\bibinfo {author} {\bibfnamefont {P.~K.}\ \bibnamefont
  {Kovtun}},\ }\bibfield  {title} {\bibinfo {title} {Lectures on hydrodynamic
  fluctuations in relativistic theories},\ }\bibfield  {booktitle} {\emph
  {\bibinfo {booktitle} {INT Summer School on Applications of String Theory
  Seattle, Washington, USA, July 18-29, 2011}},\ }\href
  {https://doi.org/10.1088/1751-8113/45/47/473001} {\bibfield  {journal}
  {\bibinfo  {journal} {J. Phys.}\ }\textbf {\bibinfo {volume} {A45}},\
  \bibinfo {pages} {473001} (\bibinfo {year} {2012})},\ \Eprint
  {https://arxiv.org/abs/1205.5040} {arXiv:1205.5040 [hep-th]} \BibitemShut
  {NoStop}%
\bibitem [{\citenamefont {Denicol}\ \emph {et~al.}(2011)\citenamefont
  {Denicol}, \citenamefont {Noronha}, \citenamefont {Niemi},\ and\
  \citenamefont {Rischke}}]{Denicol:2011fa}%
  \BibitemOpen
  \bibfield  {author} {\bibinfo {author} {\bibfnamefont {G.~S.}\ \bibnamefont
  {Denicol}}, \bibinfo {author} {\bibfnamefont {J.}~\bibnamefont {Noronha}},
  \bibinfo {author} {\bibfnamefont {H.}~\bibnamefont {Niemi}},\ and\ \bibinfo
  {author} {\bibfnamefont {D.~H.}\ \bibnamefont {Rischke}},\ }\bibfield
  {title} {\bibinfo {title} {Origin of the relaxation time in dissipative fluid
  dynamics},\ }\href {https://doi.org/10.1103/PhysRevD.83.074019} {\bibfield
  {journal} {\bibinfo  {journal} {Phys. Rev.}\ }\textbf {\bibinfo {volume}
  {D83}},\ \bibinfo {pages} {074019} (\bibinfo {year} {2011})},\ \Eprint
  {https://arxiv.org/abs/1102.4780} {arXiv:1102.4780 [hep-th]} \BibitemShut
  {NoStop}%
\bibitem [{\citenamefont {Denicol}\ \emph {et~al.}(2012)\citenamefont
  {Denicol}, \citenamefont {Niemi}, \citenamefont {Molnar},\ and\ \citenamefont
  {Rischke}}]{Denicol:2012cn}%
  \BibitemOpen
  \bibfield  {author} {\bibinfo {author} {\bibfnamefont {G.~S.}\ \bibnamefont
  {Denicol}}, \bibinfo {author} {\bibfnamefont {H.}~\bibnamefont {Niemi}},
  \bibinfo {author} {\bibfnamefont {E.}~\bibnamefont {Molnar}},\ and\ \bibinfo
  {author} {\bibfnamefont {D.~H.}\ \bibnamefont {Rischke}},\ }\bibfield
  {title} {\bibinfo {title} {Derivation of transient relativistic fluid
  dynamics from the {B}oltzmann equation},\ }\href
  {https://doi.org/10.1103/PhysRevD.85.114047, 10.1103/PhysRevD.91.039902}
  {\bibfield  {journal} {\bibinfo  {journal} {Phys. Rev.}\ }\textbf {\bibinfo
  {volume} {D85}},\ \bibinfo {pages} {114047} (\bibinfo {year} {2012})},\
  \bibinfo {note} {[Erratum: Phys. Rev.D91,no.3,039902(2015)]},\ \Eprint
  {https://arxiv.org/abs/1202.4551} {arXiv:1202.4551 [nucl-th]} \BibitemShut
  {NoStop}%
\bibitem [{\citenamefont {Baier}\ \emph {et~al.}(2008)\citenamefont {Baier},
  \citenamefont {Romatschke}, \citenamefont {Son}, \citenamefont {Starinets},\
  and\ \citenamefont {Stephanov}}]{Baier:2007ix}%
  \BibitemOpen
  \bibfield  {author} {\bibinfo {author} {\bibfnamefont {R.}~\bibnamefont
  {Baier}}, \bibinfo {author} {\bibfnamefont {P.}~\bibnamefont {Romatschke}},
  \bibinfo {author} {\bibfnamefont {D.~T.}\ \bibnamefont {Son}}, \bibinfo
  {author} {\bibfnamefont {A.~O.}\ \bibnamefont {Starinets}},\ and\ \bibinfo
  {author} {\bibfnamefont {M.~A.}\ \bibnamefont {Stephanov}},\ }\bibfield
  {title} {\bibinfo {title} {Relativistic viscous hydrodynamics, conformal
  invariance, and holography},\ }\href
  {https://doi.org/10.1088/1126-6708/2008/04/100} {\bibfield  {journal}
  {\bibinfo  {journal} {JHEP}\ }\textbf {\bibinfo {volume} {04}},\ \bibinfo
  {pages} {100}},\ \Eprint {https://arxiv.org/abs/0712.2451} {arXiv:0712.2451
  [hep-th]} \BibitemShut {NoStop}%
\bibitem [{\citenamefont {Moore}(2018)}]{Moore:2018mma}%
  \BibitemOpen
  \bibfield  {author} {\bibinfo {author} {\bibfnamefont {G.~D.}\ \bibnamefont
  {Moore}},\ }\bibfield  {title} {\bibinfo {title} {{Stress-stress correlator
  in $\phi^{4}$ theory: poles or a cut?}},\ }\href
  {https://doi.org/10.1007/JHEP05(2018)084} {\bibfield  {journal} {\bibinfo
  {journal} {JHEP}\ }\textbf {\bibinfo {volume} {05}},\ \bibinfo {pages}
  {084}},\ \Eprint {https://arxiv.org/abs/1803.00736} {arXiv:1803.00736
  [hep-ph]} \BibitemShut {NoStop}%
\bibitem [{\citenamefont {Ochsenfeld}\ and\ \citenamefont
  {Schlichting}(2023)}]{Ochsenfeld:2023wxz}%
  \BibitemOpen
  \bibfield  {author} {\bibinfo {author} {\bibfnamefont {S.}~\bibnamefont
  {Ochsenfeld}}\ and\ \bibinfo {author} {\bibfnamefont {S.}~\bibnamefont
  {Schlichting}},\ }\bibfield  {title} {\bibinfo {title} {{Hydrodynamic and
  non-hydrodynamic excitations in kinetic theory \textemdash{} a numerical
  analysis in scalar field theory}},\ }\href
  {https://doi.org/10.1007/JHEP09(2023)186} {\bibfield  {journal} {\bibinfo
  {journal} {JHEP}\ }\textbf {\bibinfo {volume} {09}},\ \bibinfo {pages}
  {186}},\ \Eprint {https://arxiv.org/abs/2308.04491} {arXiv:2308.04491
  [hep-th]} \BibitemShut {NoStop}%
\bibitem [{\citenamefont {Rocha}\ \emph
  {et~al.}(2024{\natexlab{b}})\citenamefont {Rocha}, \citenamefont {Danhoni},
  \citenamefont {Ingles}, \citenamefont {Denicol},\ and\ \citenamefont
  {Noronha}}]{Rocha:2024cge}%
  \BibitemOpen
  \bibfield  {author} {\bibinfo {author} {\bibfnamefont {G.~S.}\ \bibnamefont
  {Rocha}}, \bibinfo {author} {\bibfnamefont {I.}~\bibnamefont {Danhoni}},
  \bibinfo {author} {\bibfnamefont {K.}~\bibnamefont {Ingles}}, \bibinfo
  {author} {\bibfnamefont {G.~S.}\ \bibnamefont {Denicol}},\ and\ \bibinfo
  {author} {\bibfnamefont {J.}~\bibnamefont {Noronha}},\ }\bibfield  {title}
  {\bibinfo {title} {{Branch-cut in the shear-stress response function of
  massless $\lambda \varphi^4$ with Boltzmann statistics}},\ }\href@noop {} {\
  (\bibinfo {year} {2024}{\natexlab{b}})},\ \Eprint
  {https://arxiv.org/abs/2404.04679} {arXiv:2404.04679 [nucl-th]} \BibitemShut
  {NoStop}%
\bibitem [{\citenamefont {Gavassino}(2024)}]{Gavassino:2024rck}%
  \BibitemOpen
  \bibfield  {author} {\bibinfo {author} {\bibfnamefont {L.}~\bibnamefont
  {Gavassino}},\ }\bibfield  {title} {\bibinfo {title} {{Gapless
  non-hydrodynamic modes in relativistic kinetic theory}},\ }\href@noop {} {\
  (\bibinfo {year} {2024})},\ \Eprint {https://arxiv.org/abs/2404.12327}
  {arXiv:2404.12327 [nucl-th]} \BibitemShut {NoStop}%
\bibitem [{\citenamefont {Moore}\ and\ \citenamefont
  {Sohrabi}(2011)}]{Moore:2010bu}%
  \BibitemOpen
  \bibfield  {author} {\bibinfo {author} {\bibfnamefont {G.~D.}\ \bibnamefont
  {Moore}}\ and\ \bibinfo {author} {\bibfnamefont {K.~A.}\ \bibnamefont
  {Sohrabi}},\ }\bibfield  {title} {\bibinfo {title} {{Kubo Formulae for
  Second-Order Hydrodynamic Coefficients}},\ }\href
  {https://doi.org/10.1103/PhysRevLett.106.122302} {\bibfield  {journal}
  {\bibinfo  {journal} {Phys. Rev. Lett.}\ }\textbf {\bibinfo {volume} {106}},\
  \bibinfo {pages} {122302} (\bibinfo {year} {2011})},\ \Eprint
  {https://arxiv.org/abs/1007.5333} {arXiv:1007.5333 [hep-ph]} \BibitemShut
  {NoStop}%
\bibitem [{\citenamefont {Czajka}\ and\ \citenamefont
  {Jeon}(2017)}]{Czajka:2017bod}%
  \BibitemOpen
  \bibfield  {author} {\bibinfo {author} {\bibfnamefont {A.}~\bibnamefont
  {Czajka}}\ and\ \bibinfo {author} {\bibfnamefont {S.}~\bibnamefont {Jeon}},\
  }\bibfield  {title} {\bibinfo {title} {Kubo formulas for the shear and bulk
  viscosity relaxation times and the scalar field theory shear $\tau_\pi$
  calculation},\ }\href {https://doi.org/10.1103/PhysRevC.95.064906} {\bibfield
   {journal} {\bibinfo  {journal} {Phys. Rev.}\ }\textbf {\bibinfo {volume}
  {C95}},\ \bibinfo {pages} {064906} (\bibinfo {year} {2017})},\ \Eprint
  {https://arxiv.org/abs/1701.07580} {arXiv:1701.07580 [nucl-th]} \BibitemShut
  {NoStop}%
\bibitem [{\citenamefont {Schaefer}(2014)}]{Schaefer:2014aia}%
  \BibitemOpen
  \bibfield  {author} {\bibinfo {author} {\bibfnamefont {T.}~\bibnamefont
  {Schaefer}},\ }\bibfield  {title} {\bibinfo {title} {{Viscosity spectral
  function of a scale invariant nonrelativistic fluid from holography}},\
  }\href {https://doi.org/10.1103/PhysRevD.90.106008} {\bibfield  {journal}
  {\bibinfo  {journal} {Phys. Rev. D}\ }\textbf {\bibinfo {volume} {90}},\
  \bibinfo {pages} {106008} (\bibinfo {year} {2014})},\ \Eprint
  {https://arxiv.org/abs/1408.4503} {arXiv:1408.4503 [hep-th]} \BibitemShut
  {NoStop}%
\bibitem [{\citenamefont {{De Schepper}}\ \emph {et~al.}(1974)\citenamefont
  {{De Schepper}}, \citenamefont {{Van Beyeren}},\ and\ \citenamefont
  {Ernst}}]{DESCHEPPER19741}%
  \BibitemOpen
  \bibfield  {author} {\bibinfo {author} {\bibfnamefont {I.}~\bibnamefont {{De
  Schepper}}}, \bibinfo {author} {\bibfnamefont {H.}~\bibnamefont {{Van
  Beyeren}}},\ and\ \bibinfo {author} {\bibfnamefont {M.}~\bibnamefont
  {Ernst}},\ }\bibfield  {title} {\bibinfo {title} {The nonexistence of the
  linear diffusion equation beyond fick's law},\ }\href
  {https://doi.org/https://doi.org/10.1016/0031-8914(74)90290-0} {\bibfield
  {journal} {\bibinfo  {journal} {Physica}\ }\textbf {\bibinfo {volume} {75}},\
  \bibinfo {pages} {1} (\bibinfo {year} {1974})}\BibitemShut {NoStop}%
\bibitem [{\citenamefont {Kovtun}\ and\ \citenamefont
  {Yaffe}(2003)}]{Kovtun:2003vj}%
  \BibitemOpen
  \bibfield  {author} {\bibinfo {author} {\bibfnamefont {P.}~\bibnamefont
  {Kovtun}}\ and\ \bibinfo {author} {\bibfnamefont {L.~G.}\ \bibnamefont
  {Yaffe}},\ }\bibfield  {title} {\bibinfo {title} {{Hydrodynamic fluctuations,
  long time tails, and supersymmetry}},\ }\href
  {https://doi.org/10.1103/PhysRevD.68.025007} {\bibfield  {journal} {\bibinfo
  {journal} {Phys. Rev. D}\ }\textbf {\bibinfo {volume} {68}},\ \bibinfo
  {pages} {025007} (\bibinfo {year} {2003})},\ \Eprint
  {https://arxiv.org/abs/hep-th/0303010} {arXiv:hep-th/0303010} \BibitemShut
  {NoStop}%
\bibitem [{\citenamefont {Kovtun}\ \emph {et~al.}(2011)\citenamefont {Kovtun},
  \citenamefont {Moore},\ and\ \citenamefont {Romatschke}}]{Kovtun:2011np}%
  \BibitemOpen
  \bibfield  {author} {\bibinfo {author} {\bibfnamefont {P.}~\bibnamefont
  {Kovtun}}, \bibinfo {author} {\bibfnamefont {G.~D.}\ \bibnamefont {Moore}},\
  and\ \bibinfo {author} {\bibfnamefont {P.}~\bibnamefont {Romatschke}},\
  }\bibfield  {title} {\bibinfo {title} {{The stickiness of sound: An absolute
  lower limit on viscosity and the breakdown of second order relativistic
  hydrodynamics}},\ }\href {https://doi.org/10.1103/PhysRevD.84.025006}
  {\bibfield  {journal} {\bibinfo  {journal} {Phys. Rev. D}\ }\textbf {\bibinfo
  {volume} {84}},\ \bibinfo {pages} {025006} (\bibinfo {year} {2011})},\
  \Eprint {https://arxiv.org/abs/1104.1586} {arXiv:1104.1586 [hep-ph]}
  \BibitemShut {NoStop}%
\bibitem [{\citenamefont {Akamatsu}\ \emph {et~al.}(2017)\citenamefont
  {Akamatsu}, \citenamefont {Mazeliauskas},\ and\ \citenamefont
  {Teaney}}]{Akamatsu:2016llw}%
  \BibitemOpen
  \bibfield  {author} {\bibinfo {author} {\bibfnamefont {Y.}~\bibnamefont
  {Akamatsu}}, \bibinfo {author} {\bibfnamefont {A.}~\bibnamefont
  {Mazeliauskas}},\ and\ \bibinfo {author} {\bibfnamefont {D.}~\bibnamefont
  {Teaney}},\ }\bibfield  {title} {\bibinfo {title} {{A kinetic regime of
  hydrodynamic fluctuations and long time tails for a Bjorken expansion}},\
  }\href {https://doi.org/10.1103/PhysRevC.95.014909} {\bibfield  {journal}
  {\bibinfo  {journal} {Phys. Rev. C}\ }\textbf {\bibinfo {volume} {95}},\
  \bibinfo {pages} {014909} (\bibinfo {year} {2017})},\ \Eprint
  {https://arxiv.org/abs/1606.07742} {arXiv:1606.07742 [nucl-th]} \BibitemShut
  {NoStop}%
\bibitem [{\citenamefont {Mullins}\ \emph {et~al.}(2023)\citenamefont
  {Mullins}, \citenamefont {Hippert},\ and\ \citenamefont
  {Noronha}}]{Mullins:2023tjg}%
  \BibitemOpen
  \bibfield  {author} {\bibinfo {author} {\bibfnamefont {N.}~\bibnamefont
  {Mullins}}, \bibinfo {author} {\bibfnamefont {M.}~\bibnamefont {Hippert}},\
  and\ \bibinfo {author} {\bibfnamefont {J.}~\bibnamefont {Noronha}},\
  }\bibfield  {title} {\bibinfo {title} {{Stochastic fluctuations in
  relativistic fluids: Causality, stability, and the information current}},\
  }\href {https://doi.org/10.1103/PhysRevD.108.076013} {\bibfield  {journal}
  {\bibinfo  {journal} {Phys. Rev. D}\ }\textbf {\bibinfo {volume} {108}},\
  \bibinfo {pages} {076013} (\bibinfo {year} {2023})},\ \Eprint
  {https://arxiv.org/abs/2306.08635} {arXiv:2306.08635 [nucl-th]} \BibitemShut
  {NoStop}%
\bibitem [{\citenamefont {Soares~Rocha}\ \emph {et~al.}(2024)\citenamefont
  {Soares~Rocha}, \citenamefont {Gavassino},\ and\ \citenamefont
  {Mullins}}]{SoaresRocha:2024afv}%
  \BibitemOpen
  \bibfield  {author} {\bibinfo {author} {\bibfnamefont {G.}~\bibnamefont
  {Soares~Rocha}}, \bibinfo {author} {\bibfnamefont {L.}~\bibnamefont
  {Gavassino}},\ and\ \bibinfo {author} {\bibfnamefont {N.}~\bibnamefont
  {Mullins}},\ }\bibfield  {title} {\bibinfo {title} {{Modelling stochastic
  fluctuations in relativistic kinetic theory}},\ }\href@noop {} {\  (\bibinfo
  {year} {2024})},\ \Eprint {https://arxiv.org/abs/2405.10878}
  {arXiv:2405.10878 [nucl-th]} \BibitemShut {NoStop}%
\bibitem [{\citenamefont {Israel}(1976{\natexlab{b}})}]{MIS-2}%
  \BibitemOpen
  \bibfield  {author} {\bibinfo {author} {\bibfnamefont {W.}~\bibnamefont
  {Israel}},\ }\bibfield  {title} {\bibinfo {title} {Nonstationary irreversible
  thermodynamics: A causal relativistic theory},\ }\href@noop {} {\bibfield
  {journal} {\bibinfo  {journal} {Ann. Phys.}\ }\textbf {\bibinfo {volume}
  {100}},\ \bibinfo {pages} {310} (\bibinfo {year}
  {1976}{\natexlab{b}})}\BibitemShut {NoStop}%
\bibitem [{\citenamefont {Olson}(1990)}]{Olson:1989ey}%
  \BibitemOpen
  \bibfield  {author} {\bibinfo {author} {\bibfnamefont {T.~S.}\ \bibnamefont
  {Olson}},\ }\bibfield  {title} {\bibinfo {title} {Stability and causality in
  the {I}srael-{S}tewart energy frame theory},\ }\href
  {https://doi.org/10.1016/0003-4916(90)90366-V} {\bibfield  {journal}
  {\bibinfo  {journal} {Annals Phys.}\ }\textbf {\bibinfo {volume} {199}},\
  \bibinfo {pages} {18} (\bibinfo {year} {1990})}\BibitemShut {NoStop}%
\bibitem [{\citenamefont {Brito}\ and\ \citenamefont
  {Denicol}(2020)}]{Brito:2020nou}%
  \BibitemOpen
  \bibfield  {author} {\bibinfo {author} {\bibfnamefont {C.~V.}\ \bibnamefont
  {Brito}}\ and\ \bibinfo {author} {\bibfnamefont {G.~S.}\ \bibnamefont
  {Denicol}},\ }\bibfield  {title} {\bibinfo {title} {{Linear stability of
  Israel-Stewart theory in the presence of net-charge diffusion}},\ }\href
  {https://doi.org/10.1103/PhysRevD.102.116009} {\bibfield  {journal} {\bibinfo
   {journal} {Phys. Rev. D}\ }\textbf {\bibinfo {volume} {102}},\ \bibinfo
  {pages} {116009} (\bibinfo {year} {2020})},\ \Eprint
  {https://arxiv.org/abs/2007.16141} {arXiv:2007.16141 [nucl-th]} \BibitemShut
  {NoStop}%
\bibitem [{\citenamefont {Bemfica}\ \emph {et~al.}(2020)\citenamefont
  {Bemfica}, \citenamefont {Disconzi}, \citenamefont {Hoang}, \citenamefont
  {Noronha},\ and\ \citenamefont {Radosz}}]{Bemfica:2020xym}%
  \BibitemOpen
  \bibfield  {author} {\bibinfo {author} {\bibfnamefont {F.~S.}\ \bibnamefont
  {Bemfica}}, \bibinfo {author} {\bibfnamefont {M.~M.}\ \bibnamefont
  {Disconzi}}, \bibinfo {author} {\bibfnamefont {V.}~\bibnamefont {Hoang}},
  \bibinfo {author} {\bibfnamefont {J.}~\bibnamefont {Noronha}},\ and\ \bibinfo
  {author} {\bibfnamefont {M.}~\bibnamefont {Radosz}},\ }\bibfield  {title}
  {\bibinfo {title} {{Nonlinear Constraints on Relativistic Fluids Far From
  Equilibrium}},\ }\href@noop {} {\  (\bibinfo {year} {2020})},\ \Eprint
  {https://arxiv.org/abs/2005.11632} {arXiv:2005.11632 [hep-th]} \BibitemShut
  {NoStop}%
\bibitem [{\citenamefont {Koide}\ \emph {et~al.}(2009)\citenamefont {Koide},
  \citenamefont {Nakano},\ and\ \citenamefont {Kodama}}]{Koide:2009sy}%
  \BibitemOpen
  \bibfield  {author} {\bibinfo {author} {\bibfnamefont {T.}~\bibnamefont
  {Koide}}, \bibinfo {author} {\bibfnamefont {E.}~\bibnamefont {Nakano}},\ and\
  \bibinfo {author} {\bibfnamefont {T.}~\bibnamefont {Kodama}},\ }\bibfield
  {title} {\bibinfo {title} {{Shear viscosity coefficient and relaxation time
  of causal dissipative hydrodynamics in QCD}},\ }\href
  {https://doi.org/10.1103/PhysRevLett.103.052301} {\bibfield  {journal}
  {\bibinfo  {journal} {Phys. Rev. Lett.}\ }\textbf {\bibinfo {volume} {103}},\
  \bibinfo {pages} {052301} (\bibinfo {year} {2009})},\ \Eprint
  {https://arxiv.org/abs/0901.3707} {arXiv:0901.3707 [hep-th]} \BibitemShut
  {NoStop}%
\bibitem [{\citenamefont {Forster}(1994)}]{forster}%
  \BibitemOpen
  \bibfield  {author} {\bibinfo {author} {\bibfnamefont {D.}~\bibnamefont
  {Forster}},\ }\href@noop {} {\emph {\bibinfo {title} {Hydrodynamic
  fluctuations, broken symmetry, and correlation functions}}}\ (\bibinfo
  {publisher} {Westview Press},\ \bibinfo {year} {1994})\ p.\ \bibinfo {pages}
  {352}\BibitemShut {NoStop}%
\bibitem [{\citenamefont {Israel}\ and\ \citenamefont {Stewart}(1976)}]{MIS-3}%
  \BibitemOpen
  \bibfield  {author} {\bibinfo {author} {\bibfnamefont {W.}~\bibnamefont
  {Israel}}\ and\ \bibinfo {author} {\bibfnamefont {J.~M.}\ \bibnamefont
  {Stewart}},\ }\bibfield  {title} {\bibinfo {title} {Thermodynamics of
  nonstationary and transient effects in a relativistic gas},\ }\href@noop {}
  {\bibfield  {journal} {\bibinfo  {journal} {Phys. Lett. A}\ }\textbf
  {\bibinfo {volume} {58}},\ \bibinfo {pages} {213} (\bibinfo {year}
  {1976})}\BibitemShut {NoStop}%
\bibitem [{\citenamefont {Stewart}(1977)}]{MIS-4}%
  \BibitemOpen
  \bibfield  {author} {\bibinfo {author} {\bibfnamefont {J.~M.}\ \bibnamefont
  {Stewart}},\ }\bibfield  {title} {\bibinfo {title} {On transient relativistic
  thermodynamics and kinetic theory},\ }\href
  {http://rspa.royalsocietypublishing.org/content/357/1688/59} {\bibfield
  {journal} {\bibinfo  {journal} {Proc. R. Soc. London, Ser. A}\ }\textbf
  {\bibinfo {volume} {357}},\ \bibinfo {pages} {59} (\bibinfo {year}
  {1977})}\BibitemShut {NoStop}%
\bibitem [{\citenamefont {Gavassino}\ \emph {et~al.}(2022)\citenamefont
  {Gavassino}, \citenamefont {Antonelli},\ and\ \citenamefont
  {Haskell}}]{Gavassino:2021kjm}%
  \BibitemOpen
  \bibfield  {author} {\bibinfo {author} {\bibfnamefont {L.}~\bibnamefont
  {Gavassino}}, \bibinfo {author} {\bibfnamefont {M.}~\bibnamefont
  {Antonelli}},\ and\ \bibinfo {author} {\bibfnamefont {B.}~\bibnamefont
  {Haskell}},\ }\bibfield  {title} {\bibinfo {title} {{Thermodynamic Stability
  Implies Causality}},\ }\href {https://doi.org/10.1103/PhysRevLett.128.010606}
  {\bibfield  {journal} {\bibinfo  {journal} {Phys. Rev. Lett.}\ }\textbf
  {\bibinfo {volume} {128}},\ \bibinfo {pages} {010606} (\bibinfo {year}
  {2022})},\ \Eprint {https://arxiv.org/abs/2105.14621} {arXiv:2105.14621
  [gr-qc]} \BibitemShut {NoStop}%
\bibitem [{\citenamefont {Ghiglieri}\ \emph {et~al.}(2018)\citenamefont
  {Ghiglieri}, \citenamefont {Moore},\ and\ \citenamefont
  {Teaney}}]{Ghiglieri:2018dgf}%
  \BibitemOpen
  \bibfield  {author} {\bibinfo {author} {\bibfnamefont {J.}~\bibnamefont
  {Ghiglieri}}, \bibinfo {author} {\bibfnamefont {G.~D.}\ \bibnamefont
  {Moore}},\ and\ \bibinfo {author} {\bibfnamefont {D.}~\bibnamefont
  {Teaney}},\ }\bibfield  {title} {\bibinfo {title} {{Second-order
  Hydrodynamics in Next-to-Leading-Order QCD}},\ }\href
  {https://doi.org/10.1103/PhysRevLett.121.052302} {\bibfield  {journal}
  {\bibinfo  {journal} {Phys. Rev. Lett.}\ }\textbf {\bibinfo {volume} {121}},\
  \bibinfo {pages} {052302} (\bibinfo {year} {2018})},\ \Eprint
  {https://arxiv.org/abs/1805.02663} {arXiv:1805.02663 [hep-ph]} \BibitemShut
  {NoStop}%
\bibitem [{\citenamefont {Romatschke}\ and\ \citenamefont
  {Son}(2009)}]{Romatschke:2009ng}%
  \BibitemOpen
  \bibfield  {author} {\bibinfo {author} {\bibfnamefont {P.}~\bibnamefont
  {Romatschke}}\ and\ \bibinfo {author} {\bibfnamefont {D.~T.}\ \bibnamefont
  {Son}},\ }\bibfield  {title} {\bibinfo {title} {{Spectral sum rules for the
  quark-gluon plasma}},\ }\href {https://doi.org/10.1103/PhysRevD.80.065021}
  {\bibfield  {journal} {\bibinfo  {journal} {Phys. Rev. D}\ }\textbf {\bibinfo
  {volume} {80}},\ \bibinfo {pages} {065021} (\bibinfo {year} {2009})},\
  \Eprint {https://arxiv.org/abs/0903.3946} {arXiv:0903.3946 [hep-ph]}
  \BibitemShut {NoStop}%
\bibitem [{\citenamefont {Meyer}(2010)}]{Meyer:2010gu}%
  \BibitemOpen
  \bibfield  {author} {\bibinfo {author} {\bibfnamefont {H.~B.}\ \bibnamefont
  {Meyer}},\ }\bibfield  {title} {\bibinfo {title} {{Lattice Gauge Theory Sum
  Rule for the Shear Channel}},\ }\href
  {https://doi.org/10.1103/PhysRevD.82.054504} {\bibfield  {journal} {\bibinfo
  {journal} {Phys. Rev. D}\ }\textbf {\bibinfo {volume} {82}},\ \bibinfo
  {pages} {054504} (\bibinfo {year} {2010})},\ \Eprint
  {https://arxiv.org/abs/1005.2686} {arXiv:1005.2686 [hep-lat]} \BibitemShut
  {NoStop}%
\bibitem [{\citenamefont {Kovtun}\ \emph {et~al.}(2005)\citenamefont {Kovtun},
  \citenamefont {Son},\ and\ \citenamefont {Starinets}}]{Kovtun:2004de}%
  \BibitemOpen
  \bibfield  {author} {\bibinfo {author} {\bibfnamefont {P.}~\bibnamefont
  {Kovtun}}, \bibinfo {author} {\bibfnamefont {D.~T.}\ \bibnamefont {Son}},\
  and\ \bibinfo {author} {\bibfnamefont {A.~O.}\ \bibnamefont {Starinets}},\
  }\bibfield  {title} {\bibinfo {title} {Viscosity in strongly interacting
  quantum field theories from black hole physics},\ }\href
  {https://doi.org/10.1103/PhysRevLett.94.111601} {\bibfield  {journal}
  {\bibinfo  {journal} {Phys. Rev. Lett.}\ }\textbf {\bibinfo {volume} {94}},\
  \bibinfo {pages} {111601} (\bibinfo {year} {2005})},\ \Eprint
  {https://arxiv.org/abs/hep-th/0405231} {arXiv:hep-th/0405231} \BibitemShut
  {NoStop}%
\bibitem [{\citenamefont {Young}\ \emph {et~al.}(2015)\citenamefont {Young},
  \citenamefont {Kapusta}, \citenamefont {Gale}, \citenamefont {Jeon},\ and\
  \citenamefont {Schenke}}]{Young:2014pka}%
  \BibitemOpen
  \bibfield  {author} {\bibinfo {author} {\bibfnamefont {C.}~\bibnamefont
  {Young}}, \bibinfo {author} {\bibfnamefont {J.~I.}\ \bibnamefont {Kapusta}},
  \bibinfo {author} {\bibfnamefont {C.}~\bibnamefont {Gale}}, \bibinfo {author}
  {\bibfnamefont {S.}~\bibnamefont {Jeon}},\ and\ \bibinfo {author}
  {\bibfnamefont {B.}~\bibnamefont {Schenke}},\ }\bibfield  {title} {\bibinfo
  {title} {{Thermally Fluctuating Second-Order Viscous Hydrodynamics and
  Heavy-Ion Collisions}},\ }\href {https://doi.org/10.1103/PhysRevC.91.044901}
  {\bibfield  {journal} {\bibinfo  {journal} {Phys. Rev. C}\ }\textbf {\bibinfo
  {volume} {91}},\ \bibinfo {pages} {044901} (\bibinfo {year} {2015})},\
  \Eprint {https://arxiv.org/abs/1407.1077} {arXiv:1407.1077 [nucl-th]}
  \BibitemShut {NoStop}%
\bibitem [{\citenamefont {Murase}(2019)}]{Murase:2019cwc}%
  \BibitemOpen
  \bibfield  {author} {\bibinfo {author} {\bibfnamefont {K.}~\bibnamefont
  {Murase}},\ }\bibfield  {title} {\bibinfo {title} {{Causal hydrodynamic
  fluctuations in non-static and inhomogeneous backgrounds}},\ }\href
  {https://doi.org/10.1016/j.aop.2019.167969} {\bibfield  {journal} {\bibinfo
  {journal} {Annals Phys.}\ }\textbf {\bibinfo {volume} {411}},\ \bibinfo
  {pages} {167969} (\bibinfo {year} {2019})},\ \Eprint
  {https://arxiv.org/abs/1904.11217} {arXiv:1904.11217 [nucl-th]} \BibitemShut
  {NoStop}%
\bibitem [{\citenamefont {Landau}\ \emph {et~al.}(1980)\citenamefont {Landau},
  \citenamefont {Lifshitz},\ and\ \citenamefont
  {Pitaevski}}]{landau1980statisticalII}%
  \BibitemOpen
  \bibfield  {author} {\bibinfo {author} {\bibfnamefont {L.}~\bibnamefont
  {Landau}}, \bibinfo {author} {\bibfnamefont {E.}~\bibnamefont {Lifshitz}},\
  and\ \bibinfo {author} {\bibfnamefont {L.~P.}\ \bibnamefont {Pitaevski}},\
  }\href {https://books.google.com/books?id=KAlRAAAAMAAJ} {\emph {\bibinfo
  {title} {Statistical Physics}}},\ Course of theoretical physics\ (\bibinfo
  {publisher} {Pergamon Press},\ \bibinfo {year} {1980})\BibitemShut {NoStop}%
\bibitem [{\citenamefont {Callen}\ and\ \citenamefont
  {Welton}(1951)}]{Callen:1951vq}%
  \BibitemOpen
  \bibfield  {author} {\bibinfo {author} {\bibfnamefont {H.~B.}\ \bibnamefont
  {Callen}}\ and\ \bibinfo {author} {\bibfnamefont {T.~A.}\ \bibnamefont
  {Welton}},\ }\bibfield  {title} {\bibinfo {title} {{Irreversibility and
  generalized noise}},\ }\href {https://doi.org/10.1103/PhysRev.83.34}
  {\bibfield  {journal} {\bibinfo  {journal} {Phys. Rev.}\ }\textbf {\bibinfo
  {volume} {83}},\ \bibinfo {pages} {34} (\bibinfo {year} {1951})}\BibitemShut
  {NoStop}%
\bibitem [{\citenamefont {Kubo}(1957)}]{Kubo:1957mj}%
  \BibitemOpen
  \bibfield  {author} {\bibinfo {author} {\bibfnamefont {R.}~\bibnamefont
  {Kubo}},\ }\bibfield  {title} {\bibinfo {title} {{Statistical mechanical
  theory of irreversible processes. 1. General theory and simple applications
  in magnetic and conduction problems}},\ }\href
  {https://doi.org/10.1143/JPSJ.12.570} {\bibfield  {journal} {\bibinfo
  {journal} {J. Phys. Soc. Jap.}\ }\textbf {\bibinfo {volume} {12}},\ \bibinfo
  {pages} {570} (\bibinfo {year} {1957})}\BibitemShut {NoStop}%
\bibitem [{\citenamefont {Abbasi}\ \emph {et~al.}(2024)\citenamefont {Abbasi},
  \citenamefont {Kaminski},\ and\ \citenamefont {Tavakol}}]{Abbasi:2022aao}%
  \BibitemOpen
  \bibfield  {author} {\bibinfo {author} {\bibfnamefont {N.}~\bibnamefont
  {Abbasi}}, \bibinfo {author} {\bibfnamefont {M.}~\bibnamefont {Kaminski}},\
  and\ \bibinfo {author} {\bibfnamefont {O.}~\bibnamefont {Tavakol}},\
  }\bibfield  {title} {\bibinfo {title} {{Theory of Nonlinear Diffusion with a
  Physical Gapped Mode}},\ }\href
  {https://doi.org/10.1103/PhysRevLett.132.131602} {\bibfield  {journal}
  {\bibinfo  {journal} {Phys. Rev. Lett.}\ }\textbf {\bibinfo {volume} {132}},\
  \bibinfo {pages} {131602} (\bibinfo {year} {2024})},\ \Eprint
  {https://arxiv.org/abs/2212.11499} {arXiv:2212.11499 [hep-th]} \BibitemShut
  {NoStop}%
\bibitem [{\citenamefont {Cercignani}\ and\ \citenamefont
  {Kremer}(2002)}]{kremer}%
  \BibitemOpen
  \bibfield  {author} {\bibinfo {author} {\bibfnamefont {C.}~\bibnamefont
  {Cercignani}}\ and\ \bibinfo {author} {\bibfnamefont {G.~M.}\ \bibnamefont
  {Kremer}},\ }\href@noop {} {\emph {\bibinfo {title} {The Relativistic
  Boltzmann Equation: Theory and Applications}}}\ (\bibinfo  {publisher}
  {Birkhauser Verlag},\ \bibinfo {address} {Basel},\ \bibinfo {year}
  {2002})\BibitemShut {NoStop}%
\bibitem [{\citenamefont {Debbasch}\ and\ \citenamefont {{van
  Leeuwen}}(2009{\natexlab{a}})}]{DEBBASCH20091079}%
  \BibitemOpen
  \bibfield  {author} {\bibinfo {author} {\bibfnamefont {F.}~\bibnamefont
  {Debbasch}}\ and\ \bibinfo {author} {\bibfnamefont {W.}~\bibnamefont {{van
  Leeuwen}}},\ }\bibfield  {title} {\bibinfo {title} {General relativistic
  boltzmann equation, i: Covariant treatment},\ }\href
  {https://doi.org/https://doi.org/10.1016/j.physa.2008.12.023} {\bibfield
  {journal} {\bibinfo  {journal} {Physica A: Statistical Mechanics and its
  Applications}\ }\textbf {\bibinfo {volume} {388}},\ \bibinfo {pages} {1079}
  (\bibinfo {year} {2009}{\natexlab{a}})}\BibitemShut {NoStop}%
\bibitem [{\citenamefont {Debbasch}\ and\ \citenamefont {{van
  Leeuwen}}(2009{\natexlab{b}})}]{DEBBASCH20091818}%
  \BibitemOpen
  \bibfield  {author} {\bibinfo {author} {\bibfnamefont {F.}~\bibnamefont
  {Debbasch}}\ and\ \bibinfo {author} {\bibfnamefont {W.}~\bibnamefont {{van
  Leeuwen}}},\ }\bibfield  {title} {\bibinfo {title} {General relativistic
  boltzmann equation, ii: Manifestly covariant treatment},\ }\href
  {https://doi.org/https://doi.org/10.1016/j.physa.2009.01.009} {\bibfield
  {journal} {\bibinfo  {journal} {Physica A: Statistical Mechanics and its
  Applications}\ }\textbf {\bibinfo {volume} {388}},\ \bibinfo {pages} {1818}
  (\bibinfo {year} {2009}{\natexlab{b}})}\BibitemShut {NoStop}%
\bibitem [{\citenamefont {Denicol}\ \emph
  {et~al.}(2014{\natexlab{a}})\citenamefont {Denicol}, \citenamefont {Heinz},
  \citenamefont {Martinez}, \citenamefont {Noronha},\ and\ \citenamefont
  {Strickland}}]{Denicol:2014xca}%
  \BibitemOpen
  \bibfield  {author} {\bibinfo {author} {\bibfnamefont {G.~S.}\ \bibnamefont
  {Denicol}}, \bibinfo {author} {\bibfnamefont {U.~W.}\ \bibnamefont {Heinz}},
  \bibinfo {author} {\bibfnamefont {M.}~\bibnamefont {Martinez}}, \bibinfo
  {author} {\bibfnamefont {J.}~\bibnamefont {Noronha}},\ and\ \bibinfo {author}
  {\bibfnamefont {M.}~\bibnamefont {Strickland}},\ }\bibfield  {title}
  {\bibinfo {title} {New exact solution of the relativistic boltzmann equation
  and its hydrodynamic limit},\ }\href
  {https://doi.org/10.1103/PhysRevLett.113.202301} {\bibfield  {journal}
  {\bibinfo  {journal} {Phys. Rev. Lett.}\ }\textbf {\bibinfo {volume} {113}},\
  \bibinfo {pages} {202301} (\bibinfo {year} {2014}{\natexlab{a}})},\ \Eprint
  {https://arxiv.org/abs/1408.5646} {arXiv:1408.5646 [hep-ph]} \BibitemShut
  {NoStop}%
\bibitem [{\citenamefont {Denicol}\ \emph
  {et~al.}(2014{\natexlab{b}})\citenamefont {Denicol}, \citenamefont {Heinz},
  \citenamefont {Martinez}, \citenamefont {Noronha},\ and\ \citenamefont
  {Strickland}}]{Denicol:2014tha}%
  \BibitemOpen
  \bibfield  {author} {\bibinfo {author} {\bibfnamefont {G.~S.}\ \bibnamefont
  {Denicol}}, \bibinfo {author} {\bibfnamefont {U.~W.}\ \bibnamefont {Heinz}},
  \bibinfo {author} {\bibfnamefont {M.}~\bibnamefont {Martinez}}, \bibinfo
  {author} {\bibfnamefont {J.}~\bibnamefont {Noronha}},\ and\ \bibinfo {author}
  {\bibfnamefont {M.}~\bibnamefont {Strickland}},\ }\bibfield  {title}
  {\bibinfo {title} {Studying the validity of relativistic hydrodynamics with a
  new exact solution of the boltzmann equation},\ }\href
  {https://doi.org/10.1103/PhysRevD.90.125026} {\bibfield  {journal} {\bibinfo
  {journal} {Phys. Rev.}\ }\textbf {\bibinfo {volume} {D90}},\ \bibinfo {pages}
  {125026} (\bibinfo {year} {2014}{\natexlab{b}})},\ \Eprint
  {https://arxiv.org/abs/1408.7048} {arXiv:1408.7048 [hep-ph]} \BibitemShut
  {NoStop}%
\bibitem [{\citenamefont {Denicol}\ and\ \citenamefont
  {Noronha}(2024)}]{Denicol:2022bsq}%
  \BibitemOpen
  \bibfield  {author} {\bibinfo {author} {\bibfnamefont {G.~S.}\ \bibnamefont
  {Denicol}}\ and\ \bibinfo {author} {\bibfnamefont {J.}~\bibnamefont
  {Noronha}},\ }\bibfield  {title} {\bibinfo {title} {{Spectrum of the
  Boltzmann collision operator for $\lambda\phi^4$ theory in the classical
  regime}},\ }\href {https://doi.org/10.1016/j.physletb.2024.138487} {\bibfield
   {journal} {\bibinfo  {journal} {Phys. Lett. B}\ }\textbf {\bibinfo {volume}
  {850}},\ \bibinfo {pages} {138487} (\bibinfo {year} {2024})},\ \Eprint
  {https://arxiv.org/abs/2209.10370} {arXiv:2209.10370 [nucl-th]} \BibitemShut
  {NoStop}%
\end{thebibliography}%

\end{document}